%% file: paper.tex
\documentclass[letterpaper,fleqn,useAMS,usenatbib,usegraphicx]{mn2e}
\usepackage[totalwidth=480pt,totalheight=680pt,layoutvoffset=0.5cm]{geometry}
\usepackage{times}
\usepackage{amsmath}
\usepackage{amssymb}
\usepackage{flushend}
\input{ads_defns}
\newcommand{\masyr}{mas~yr$^{-1}$}
\newcommand{\kms}{\textrm{km~s$^{-1}$}}
\newcommand{\vsini}{\ensuremath{v\sin i}}

\newcommand{\rxjstar}{RX J1123.2$-$7924}

\begin{document}
\title[New low-mass members and the age of Octans]{New low-mass members of the Octans stellar association and an updated 30--40 Myr lithium age}
\author[S. J. Murphy \& W. A. Lawson]{Simon~J.~Murphy$^{1,2}$\thanks{E-mail: murphy@ari.uni-heidelberg.de} and Warrick~A.~Lawson$^2$ \\
$^1$ Gliese Fellow, Astronomisches Rechen-Institut, Zentrum f\"{u}r Astronomie der Universit\"{a}t Heidelberg, D-69120 Heidelberg, Germany\\
$^2$  School of Physical, Environmental and Mathematical Sciences, University of New South Wales Canberra, ACT 2600, Australia}

\maketitle
\begin{abstract}
The Octans association is one of several young stellar moving groups recently discovered in the Solar neighbourhood and hence a valuable laboratory for studies of stellar, circumstellar disc and planetary evolution. However, a lack of low-mass members or any members with trigonometric parallaxes means the age, distance and space motion of the group are poorly constrained. To better determine its membership and age, we present the first spectroscopic survey for new K and M-type Octans members, resulting in the discovery of 29 UV-bright K5--M4 stars with kinematics, photometry and distances consistent with existing members. Nine new members possess strong Li\,\textsc{i} $\lambda$6708 absorption, which allow us to estimate a lithium age of 30--40~Myr, similar to that of the Tucana-Horologium association and bracketed by the firm lithium depletion boundary ages of the $\beta$ Pictoris (20~Myr) and Argus/IC 2391 (50~Myr) associations. Several stars also show hints in our medium-resolution spectra of fast rotation or spectroscopic binarity. More so than other nearby associations, Octans is much larger than its age and internal velocity dispersion imply. It may be the dispersing remnant of a sparse, extended structure which includes some younger members of the  foreground Octans-Near association recently proposed by Zuckerman and collaborators.
\end{abstract}

\section{Introduction}\label{sec:introduction}

The realisation, nearly 20 years ago, that the Solar neighbourhood is home to a handful of stellar associations and moving groups with ages less than the Pleiades was a watershed moment in  studies of early stellar evolution, circumstellar discs and planet formation, among other topics \citep[see reviews by][]{Zuckerman04a,Torres08}. These young associations, including well-known groups around the bright stars $\eta$~Chamaeleontis \citep[$<$10~Myr;][]{Murphy13}, TW Hydrae \citep[$\sim$10~Myr;][]{Weinberger13}, $\beta$ Pictoris \cite[$\sim$20~Myr;][]{Mamajek14} and AB Doradus \citep[$\sim$120~Myr;][]{Barenfeld13} are the kinematically coherent remnants of recent low-density star formation \citep{Mamajek01,Ortega09} and bridge the divide between rich star-forming regions and the dispersed, older field population. Because they are some of the best places to observe circumstellar discs and nascent planetary systems at high angular resolution and sensitivity \citep{Chauvin04,Marois08,Sicilia-Aguilar09}, the study of such associations to date has understandably concentrated on those nearby groups accessible by the \emph{Hipparcos} astrometric satellite.  Thanks to their youth ($\lesssim$100~Myr) and proximity ($\lesssim$100~pc), it is possible to observe the mass function in nearby groups down to low and substellar masses \citep[e.g.][]{Kraus14,Gagne14}. Moreover, the availability of accurate trigonometric parallaxes and proper motions from \emph{Hipparcos} (for at least the higher mass members) allows the calculation of the mean group space motion, from which kinematic distances can be easily derived for fainter members lacking parallaxes. 

In contrast to its closer cousins, the Octans association is not as well characterised. This is due to its deep southern declination, larger mean distance \citep[$d_{\textrm{kin}}=141\pm34$~pc;][]{Torres08} and subsequent absence of any members with \emph{Hipparcos} observations. First proposed in the SACY (Search for Associations Containing Young stars) programme of \citet{Torres03b,Torres03}, Octans originally comprised six young FGK-type stars around the south celestial pole. The review of \citet{Torres08} increased its membership to 15 active solar-type stars, all detected in X-rays by the \emph{ROSAT} satellite.  \citeauthor{Torres08} used a convergence method (see \S\ref{sec:kinematicmembers}) and an assumed age of 10~Myr to derive a heliocentric velocity and kinematic distance for each member . A young age was supported by the lack of lithium depletion and fast rotation rates ($\vsini=20$--200~\kms) observed in the group. Their membership solution has a space motion very different from other young groups and is much more extended than Octans' age and observed internal velocity dispersion would suggest. As a compromise, they proposed an a posteriori age of about 20~Myr, but this is not well constrained by the known members. In addition to being a fundamental stellar parameter, age is vital for understanding the evolution of circumstellar material and the masses of any hitherto undiscovered exoplanets. Determining a better age estimate for Octans is therefore a key goal of our present work.

Notably, there are no confirmed Octans members later than spectral type K1, despite the fact that these stars should outnumber the known solar-type population nearly ten-fold  \citep{Kroupa02}. This dearth of low-mass stars is unsurprising given the early reliance on \emph{ROSAT} detections to define the group. Assuming saturated X-ray emission typical in young stars ($\log L_{X}/L_{\textrm{bol}}\approx-3$), a 10--20~Myr-old star at 140~pc would need to be of K3 spectral type or earlier to have been detected by \emph{ROSAT}. As well as being more populous, these putative low-mass members are vital for better estimating Octans' age. Photospheric lithium depletion (measured by the equivalent width of the unresolved Li\,\textsc{i} $\lambda$6708 doublet) can be used as a distance-independent clock at pre-main sequence ages, and is particularly effective for relative age rankings \citep{Mentuch08,Soderblom13}. However, because of their shallow convective zones, the slow lithium depletion observed in young solar-type stars can only constrain the existing Octans membership to approximately Pleiades age or younger. By contrast, lithium depletion is very sensitive to both age and stellar mass at K and M-type spectral types and ages of 10--100~Myr. With a large enough sample of low-mass members it should therefore be possible to derive an ensemble lithium depletion age for the group.

To enable this analysis, we present the first systematic survey for low-mass members of Octans, resulting in the confirmation of 29 new K5--M4 members at estimated kinematic distances of 85--181~pc. In \S\ref{sec:candselection} we outline our candidate selection techniques, followed in \S\ref{sec:spectra} by a description of the ANU 2.3-m spectroscopy used to acquire radial velocities, spectral types and youth indicators. We combine these measurements with additional members proposed in the literature in \S\ref{sec:membership} to determine a self-consistent membership for Octans, and conclude in \S\ref{sec:discussion} with a brief discussion of the structure of the association and its relationship to the foreground `Octans-Near' group recently proposed by \citet{Zuckerman13}.

\section{Candidate selection}\label{sec:candselection}

We concentrate in this survey on identifying new K and M spectral type members of Octans, as these stars are expected to show age-dependent levels of lithium depletion. As the primary photometric and astrometric catalogue for this work, we adopted the Fourth Yale/San Juan Southern Proper Motion survey \citep[SPM4;][]{Girard11}. SPM4 contains absolute proper motions and $BV$ photometry \citep[with $JHK_{s}$ from 2MASS;][]{Skrutskie06} complete to $V=17.5$ for approximately 10$^{8}$ sources between $-90^{\circ}<\delta<-20^{\circ}$. Compared to the recent UCAC4 proper motion catalogue \citep{Zacharias13}, SPM4 is more complete at fainter magnitudes \citep[see discussion in][]{Murphy13} and its $BV$ photometry is more spatially complete than the APASS \citep{Henden12} DR6 photometry used in UCAC4. Agreement between SPM4 and APASS $V$ is typically better than 0.15~mag for most objects. As its photometric data come from several sources, no formal errors are given for the $B$ and $V$ magnitudes in SPM4. However, nearly two thirds of the $V$ magnitudes are derived from second-epoch CCD frames calibrated against Tycho-2 \citep{Hog00} which should be relatively homogeneous. In any case, the broad baseline of our adopted $V-K_{s}$ colour index should mitigate the effect of photometric errors. 

\subsection{Initial photometric and kinematic selection} \label{sec:kinematicselection}

We first queried SPM4 to pre-select those late-type, unreddened objects with colour-magnitude positions indicative of nearby, young stars. This was accomplished using the Virtual Observatory Table Access Protocol standard \citep[TAP;][]{Dowler11} which allows structured (ADQL) queries of catalogue data\footnote{TAP service hosted by the German Astrophysical Virtual Observatory,  VO resource identifier \texttt{ivo://org.gavo.dc/tap}. To aid reproducibility we include these identifiers (with entries in the various global VO registries) to uniquely reference the specific VO service and data used.}: 
\begin{verbatim}
  SELECT * 
  FROM spm4.main 
  WHERE magv<18
  AND magv-magk>2 
  AND magj-magh<0.7 
  AND magv<2*(magv-magk)+9
\end{verbatim}
The $V<18$~mag limit was imposed to minimise photometric and proper motion errors, and to permit spectroscopic follow-up observations on a 2-m telescope. The last clause is approximately equivalent to selecting those stars above a 50~Myr isochrone at a distance of 300~pc. This initial query resulted in $5.4\times10^{6}$ sources across the entire SPM4 survey footprint ($\delta < -20^{\circ}$). 

It is a requirement that moving group members share a common motion through space. Because Octans members span a large range of distances and occupy a substantial fraction of the sky, any kinematic selection must be made against the projection of the mean group space motion into an expected proper motion and radial velocity at the position of each candidate.  Projecting the \citet{Torres08} space motion\footnote{Following right-handed convention, we define the heliocentric position and velocity vectors $X$ and $U$ as positive towards the Galactic centre, $Y$ and $V$ positive in the direction of Galactic rotation, and $Z$ and $W$ positive towards the north Galactic pole.}, $(U,V,W)=(-14.5, -3.6, -11.2)$~\kms, over a range of distances ($5<d<300$~pc, with 5~pc increments), we retained those sources whose lowest total proper motion difference,
\begin{equation}[(\mu_\alpha\cos\delta_\textrm{obs}-\mu_\alpha\cos\delta_\textrm{expected})^{2} + (\mu_{\delta\textrm{obs}}-\mu_{\delta\textrm{expected}})^{2}]^\frac{1}{2}
\end{equation}
 was less than twice the total proper motion error, which we restricted to $\sigma_\mu<5$~\masyr. This process returned $\sim$$2\times10^{5}$ potential kinematic members, each with an estimated best kinematic distance, heliocentric position and velocity. 

\subsection{\textit{GALEX} ultraviolet selection} \label{sec:nuv}

Due to their strong magnetic fields and fast rotation rates, enhanced X-ray and ultraviolet (UV) emission is a defining characteristic of young, active stars \citep[e.g.][]{Preibisch05,Zuckerman04a}. Because of the limited sensitivity of all-sky X-ray surveys like  \emph{ROSAT}, in recent years there have been several efforts to exploit deeper UV photometry from the \emph{Galaxy Evolution Explorer} mission \citep[\emph{GALEX;}][]{Martin05} to identify nearby young stars  \citep{Findeisen10,Shkolnik11,Rodriguez11,Rodriguez13}. Although less complete than a kinematic and colour-magnitude selection alone \citep{Kraus14},  UV photometry provides an efficient way to identify active, potentially young Octans members from the large list of kinematic candidates.  

\begin{figure}
   \centering
   \includegraphics[width=0.95\linewidth]{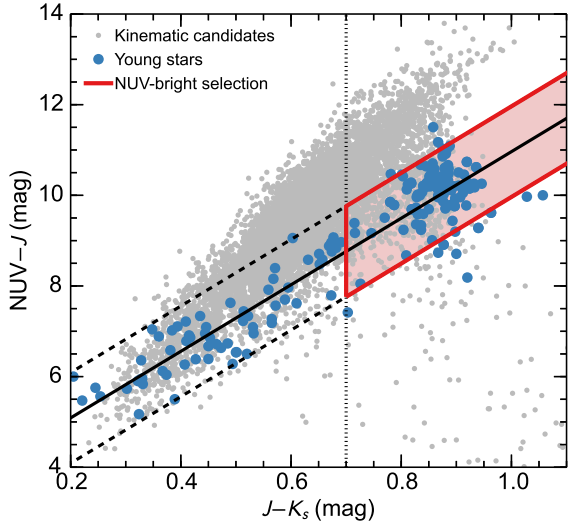} \hspace{3mm}
   \caption{\emph{GALEX} NUV selection criteria.  Known young stars (blue points) are NUV-bright compared to the majority of kinematic candidates (small grey points) at later spectral types. We select those candidates redward of $J-K_{s}=0.7$ (dotted line) which lie within 1~mag of the linear fit to the young star sample (red shaded region).}
   \label{fig:nuv}
\end{figure}

Approximately 25,000 (13 per cent) of the remaining candidates returned a match against \emph{GALEX} GR6\footnote{\texttt{ivo://wfau.roe.ac.uk/galexgr6-dsa} } within a radius of 2~arcsec. The satellite did not generally observe within 20$^{\circ}$ of the Galactic equator, hence our search will be incomplete at these latitudes. Our selection of UV-bright stars follows that of \citet{Rodriguez11}, who combined near-UV (NUV; 1800--2800~\AA) photometry and 2MASS $J$ to identify stars with  NUV$-J$ excess. For this work and the colour-magnitude selection described in the next section it is useful to define a canonical sample of young ($<$100~Myr) stars to compare against. For this we combined the young moving group memberships of \citet{Torres08} and \citet{Malo13} to collate 183 early-A to mid-M stars with NUV detections, SPM4 $V$ photometry and trigonometric or kinematic distances. Our NUV selection is illustrated in Fig.\,\ref{fig:nuv}, where the young star sample is plotted with the kinematic candidates. Moving to later spectral types, the young stars appear increasingly NUV-bright compared to the majority of the candidates, which are likely older field stars with kinematics mimicking those of Octans members. We select as NUV-bright those candidates with $J-K_{s}>0.7$ (approximately K5) that lie within 1~mag of a linear regression to the young star sample, NUV$-J=7.3(J-K_{s})+3.6$.  

\subsection{Colour-magnitude selection} \label{sec:cmdselect}

\begin{figure}
   \centering
    \includegraphics[width=0.95\linewidth]{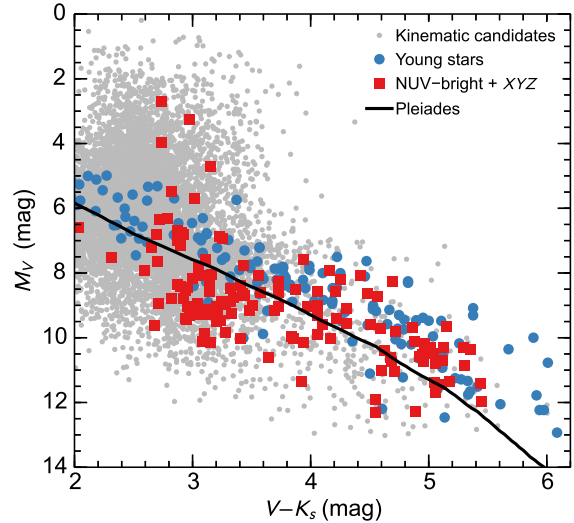} 
   \caption{Colour-magnitude selection. Only those NUV-bright candidates that lie within the volume of known Octans members are plotted (see text). We select for spectroscopic follow-up those 61 stars lying above the \citet{Stauffer07} empirical Pleiades isochrone (solid line).}
   \label{fig:cmdselect}
\end{figure}

Young low-mass stars have elevated luminosities which place them up to several magnitudes above the main sequence. It is therefore vital to confirm that the kinematic distances of the candidates agree with their luminosity distances, modulo the effects of unresolved binarity.   To further winnow the sample, we first required that the $\sim$1,600 NUV-bright candidates had heliocentric positions within the volume of known Octans members; 
\begin{align}\label{eqn:xyz}
-100<&X<150~\textrm{pc},\nonumber\\
-150<&Y<-50~\textrm{pc},\\
-100<&Z<0~\textrm{pc.}\nonumber
 \end{align}
 It is possible that Octans is larger than these limits (see \S\ref{sec:structure}). However, as with previous criteria we prioritised efficiency over completeness and retained only those 130 stars (8 per cent of the NUV-bright sample) which satisfied Eqn.\,\ref{eqn:xyz}. Fig.\,\ref{fig:cmdselect} shows the $M_{V}$ versus $V-K_{s}$ colour-magnitude diagram for these stars, compared to the empirical $\sim$100~Myr Pleiades isochrone presented by \citet{Stauffer07}. At $V-K_{s}>3.5$ the NUV-bright sample tends to lie above the Pleiades relation, in the region occupied by the moving group members, supporting the hypothesis that these stars are indeed young. Because the age of Octans remains uncertain (one of the motivations for this work) but appears to be younger than 100~Myr, we selected as spectroscopic targets the 61 candidates lying above the Pleiades isochrone. The positions and photometry of these stars are listed in Table~\ref{tab:spectroscopy}.

\section{Spectroscopic observations}\label{sec:spectra}

\begin{figure*}
   \centering
      \includegraphics[width=0.46\linewidth]{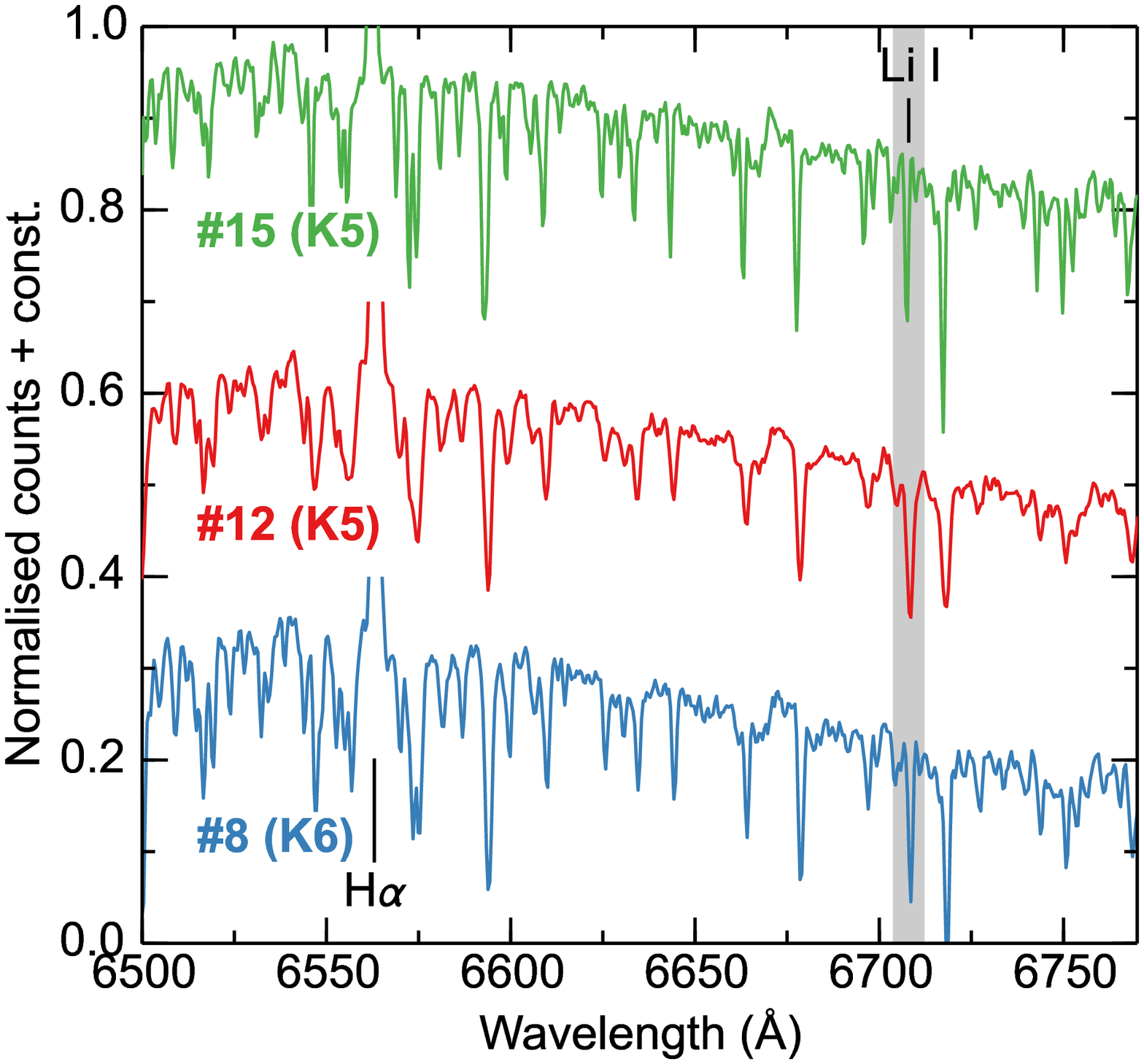} 
   \includegraphics[width=0.46\linewidth]{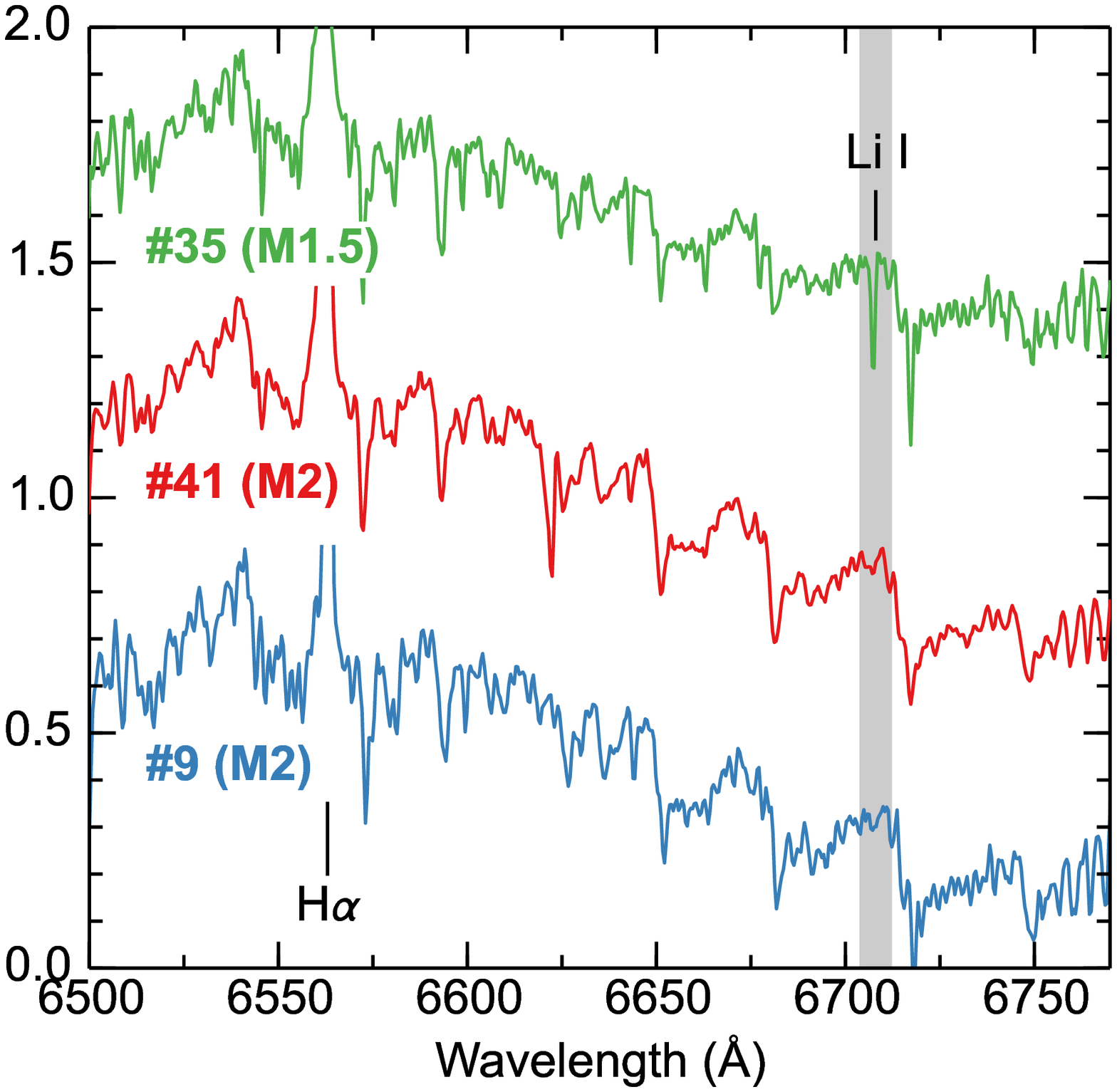} 
   \caption{H$\alpha$ region spectra of six new Octans members confirmed in this work. The full WiFeS spectra span 5300--7100~\AA. The spectra were divided by their mean flux prior to plotting and the positions of the Li\,\textsc{i}  absorption lines (not present in stars \#9 and \#41) and truncated H$\alpha$ emission lines are marked. The spectral features of star \#12, and to a lesser extent \#41, appear broader than the stars above and below them in each panel. These two stars belong to the group of candidates with wider cross-correlation functions, suggestive of fast rotation or unresolved spectroscopic binarity (see Fig.~\ref{fig:broadccf} and \S\ref{sec:broadccf}). }
   \label{fig:spectra}
\end{figure*}

The kinematic selection described in \S\ref{sec:kinematicselection} predicts a radial velocity for each candidate assuming it is a true (and non-spectroscopic binary) member of Octans. To measure these radial velocities, estimate spectral types and check for the Li\,\textsc{i} $\lambda$6708 youth indicator, we observed 57/61 candidates and a selection of K and M-type radial velocity standards in 2014 January, April and May with the Wide Field Spectrograph \citep[WiFeS;][]{Dopita07} on the ANU 2.3-m telescope at Siding Spring Observatory. Observations were made in half field (12$\times$38 arcsec) mode with 2$\times$ spatial binning (1 arcsec spaxels). The $R$7000 grating and $RT$480 (January, May) and $RT$560 (April) dichroics gave a resolution of $\lambda/\Delta\lambda\approx7000$ and wavelength coverage from 5300--7100~\AA. Exposure times were 600--5400~s (split into 3$\times$1800~s to minimise cosmic ray hits).   The WiFeS field of view was generally aligned to the parallactic angle prior to each exposure, although in three cases (\#18, \#32, \#43) the candidate was clearly elongated or a resolved double and  we rotated the field such that its long axis was perpendicular to the position angle of the components, maximising the separation of the spectra on different image slices. Each component was then extracted separately. No flux standards were observed and the spectra were not corrected for telluric absorption. 

We used custom \textsc{iraf}, \textsc{figaro} and \textsc{python} routines to rectify, extract, wavelength-calibrate and combine the 3--5 (depending on seeing) 38$\times$1~arcsec image slices that contained the majority of the stellar flux. The slices were treated like long-slit spectra and individually extracted and wavelength calibrated against NeAr arc frames taken immediately after each exposure. Typical H$\alpha$ region spectra are plotted in Fig.~\ref{fig:spectra}.\footnote{Reduced FITS spectra of all objects and machine-readable versions of Tables~\ref{tab:spectroscopy}--\ref{tab:rosat} can be found on the \textsc{figshare} service at the following DOI: \texttt{http://dx.doi.org/10.6084/m9.figshare.1230051}} We measured H$\alpha$ and Li\,\textsc{i} equivalent widths by fitting Gaussian profiles to each line. These measurements are listed in Table~\ref{tab:spectroscopy}. Uncertainties in the case of Li\,\textsc{i} were estimated from multiple fits, varying the pseudo-continuum and integration limits. In addition to H$\alpha$, many of the M-type candidates also showed weak ($\textrm{EW}<1$~\AA)  emission in He\,\textsc{i} $\lambda$6678 and $\lambda$5876. These lines are chromospheric in origin and are observed in active M-dwarfs of up to field age \citep{Gizis02}. 

\subsection{Radial velocities}

Radial velocities were measured by cross-correlation against the standards observed that month using a \textsc{python} implementation of the \textsc{fxcor} Fourier cross-correlation algorithm \citep{Tonry79}. Because only a single velocity standard was observed in 2014 April, we correlated the ten April observations against the May standards. Prior to cross-correlation, each spectrum was resampled onto a log-linear wavelength scale and normalised by subtracting a boxcar-smoothed copy and dividing by the standard deviation. The modest spectral resolution of WiFeS means it is important to extract as much velocity precision from the instrument as possible. We therefore used the radial velocity standards to compare combinations of the boxcar size, wavelength range and cosine bell apodisation fraction. The lowest rms velocity variation was found with a boxcar size of 2 pixels, 5500--6500~\AA\ and an apodisation of 30 per cent. This critically-narrow 2~pixel smoothing length acts like a high-pass Fourier filter, removing almost all low frequency structures in the spectra but preserving the line cores. The 25 and 19 standards observed in 2014 January and May, respectively, demonstrate that these parameters provide a high signal-to-noise velocity precision of $\lesssim$1~\kms. For the faintest candidates this falls to $\sim$2~\kms\ per exposure, determined from multiple observations of several objects. Mean radial velocities for each candidate are listed in Table~\ref{tab:spectroscopy}. For stars observed at several epochs this is the average velocity, weighted by the nightly standard deviations. Several spectroscopic binaries and candidates with broader cross-correlation functions are discussed in \S\ref{sec:binaries} and \S\ref{sec:broadccf}, respectively.

\subsection{Spectral types}\label{sec:sptypes}

Because the $R$7000 spectra were not flux-calibrated, traditional methods for determining spectral types via flux indices or template matching were impractical.  Despite this, we compared each candidate to the complete set of radial velocity standard observations, calculating the rms difference for every pairing. Spectral types were estimated by visual inspection of the ten best matches using the values tabled in SIMBAD \citep{Wenger00}. However, as every observation did not always fall on the same set of image slices in the WiFeS field, without flux calibration these spectral types are provisional. As validation we computed the spectral energy distribution (SED) of every candidate using the Virtual Observatory SED Analyser \citep[VOSA;][]{Bayo08} with SPM4, APASS and 2MASS photometry, as well as additional photometry from the DENIS \citep{Epchtein99} and \emph{Wide-field Infrared Survey Explorer}  \citep[\emph{WISE};][]{Wright10} surveys. We used VOSA to determine an effective temperature and surface gravity for every SED by fitting synthetic fluxes from solar metallicity BT-Settl models \citep{Allard12}, excluding from the fits the generally poor-quality SPM4 $B$ and \emph{WISE} $W4$ (22~\micron) measurements. Finally, we converted these temperatures to spectral types using the updated $T_{\textrm{eff}}$-colour-spectral type scale of \citet{Pecaut13}. This is also derived from BT-Settl models, but with fixed surface gravity, $\log(g)=4.3$. As discussed by \citeauthor{Pecaut13}, the synthetic fluxes do not strongly depend on the adopted gravity. In almost all cases the photometric and spectroscopic spectral types agreed to within a subtype. Adopted average values are given in Table~\ref{tab:spectroscopy}. To account for the incomplete coverage of the standards, mixed provenance of their SIMBAD spectral types and the 100~K model grid used by VOSA, we conservatively estimate errors of $\pm$1 subtype for all candidates.  From the agreement of the SED and spectroscopic spectral types, we conclude that none of the candidates is significantly reddened ($E(B-V)\lesssim0.1$~mag). This is consistent with their position below the Galactic plane ($|b|>13^{\circ}$), modest distances ($d<200$~pc) and recent three-dimensional extinction maps \citep{Lallement14}.
 
\subsection{Spectroscopic binaries}\label{sec:binaries}

Two candidates are confirmed spectroscopic binaries. The 2014 May~12 cross-correlation function (CCF) and spectrum of star \#14 is clearly resolved into two components with radial velocities $-$57~\kms\ and $+$54~\kms. An additional spectrum on 2014 Sep~7 showed a narrow, single-peaked CCF with radial velocity $-1.4\pm1.3$~\kms\ which we adopt in Table~\ref{tab:spectroscopy}. Assuming this is close to systemic, the method of \citet{Wilson41} gives a mass ratio of $r=1.0\pm0.1$. The composite spectral type is M4. Of the 12 candidates observed more than once, only candidate \#16 exhibited radial velocity variation outside of the expected errors; changing from $-$5.6~\kms\ to $-$65.7~\kms\ between 2014 Jan 17 and May 14. We attribute this to spectroscopic binarity. There are weak lines from a secondary visible in the May spectrum and a smaller secondary peak in the CCF at $+$90~\kms\ separation. By contrast, the January CCF is single-peaked and symmetric. We derive a mass ratio of $r=0.5\pm0.1$. The primary spectrum is well-matched to a spectral type of K3 at both epochs. 

\subsection{Stars with broad cross-correlation functions}\label{sec:broadccf}

\begin{figure}
   \centering
   \includegraphics[width=0.9\linewidth]{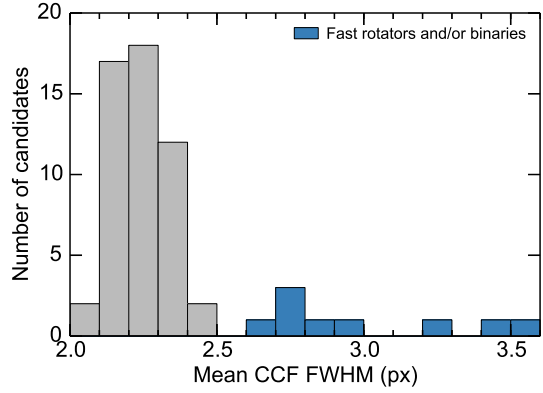}
   \caption{Distribution of mean cross-correlation function FWHM values (from 25 and 19 standards in 2014 Jan and May, respectively). Typical standard deviations around each mean value are 0.05--0.1~px. We identify as possible fast rotators or unresolved spectroscopic binaries those nine candidates with a mean FWHM $>2.5$~px (blue shading).}
   \label{fig:broadccf}
\end{figure}

Following each cross-correlation, we measured the width of the Gaussian fitted to the peak of the CCF, with the resulting mean widths plotted in Fig.~\ref{fig:broadccf}. While most candidates had a full width at half maximum (FWHM) of 2.1--2.4~px, nine stars (noted in Table~\ref{tab:spectroscopy}) had significantly larger widths and correspondingly broader spectral features (Fig.~\ref{fig:spectra}). At the modest resolution of WiFeS ($R\approx7000$; $c\Delta\lambda/\lambda\approx45$~\kms), two mechanisms can broaden spectral lines in this way; high rotation rates \citep[e.g.][]{Soderblom89} or unresolved spectroscopic binarity. We explored these effects on our spectra by convolving them with rotational kernels \citep{Gray08} and simulating equal-mass binaries of different velocity offsets. In both cases we could broaden the CCFs monotonically to the values seen in Fig.~\ref{fig:broadccf}.  For instance, we reproduced the spectrum of candidate \#12 ($\textrm{FWHM}=2.9$~px) with candidate \#15 as a narrow-lined template (Fig.~\ref{fig:spectra}) using either rotation ($\vsini\approx45$~\kms) or binarity ($\Delta \textrm{RV}\approx45$~\kms). Candidate \#12 was the only such star observed on multiple (4) epochs -- between 2014 Jan and May its mean velocity varied by only 6~\kms.  Binarity at this level would not broaden the CCF to the extent seen, so we conclude that it is likely due to rotation. Higher-resolution spectroscopy is required to separate these two effects in the other candidates.

\section{The low-mass membership of Octans}\label{sec:membership}

We can now select those candidates with kinematic, photometric and spectroscopic properties consistent with membership in a young moving group, as well as test the definition of Octans proposed by \citet{Torres08}. 

\subsection{Additional members from the literature}

Several new members have been proposed in the literature since the \citeauthor{Torres08} review. During their \emph{Spitzer Space Telescope} survey of dust around nearby F-type stars, \citet{Moor11} proposed HD~36968, which shares Octans' kinematics at a photometric distance of 140~pc and hosts a high fractional luminosity ($L_{\textrm{dust}}/L_{\textrm{bol}}\approx10^{-3}$) debris disc. \citet{Murphy13} proposed the marginal $\epsilon$~Chamaeleontis association member and spectroscopic binary \rxjstar\ as a possible member of Octans, based on a reasonable kinematic and spatial match, low Li\,\textsc{i} equivalent width and position 1.5~mag below the $\epsilon$~Cha isochrone.  Finally, from their spectroscopic follow-up of the full SACY sample, \citet{Elliott14} recently added the lithium-rich stars BD$-$20 1111 and CD$-$49 2037 and showed  HD 155177 \citep{Torres08} is a double-lined spectroscopic binary. These stars are listed with the members proposed by \citet{Torres08} in Table~\ref{tab:previousmembers} and included in the analysis below.

\subsection{Kinematic membership}\label{sec:kinematicmembers}

In previous efforts, we \citep{Murphy13} and others \citep{Torres06,Torres08,De-Silva13} have used so-called \emph{convergence} methods to simultaneously define and test the membership solutions for many nearby young groups. However, in the singular case of Octans, which lacks a good age estimate, low-mass membership or any members with trigonometric parallaxes, such an approach is problematic. Without an a priori age, velocity or distance anchor, it is difficult to self-consistently select potential members given the degeneracies between photometric (isochronal) age, kinematic distance and space motion. Nevertheless, it is convenient to use such techniques in the first instance to refine the group space motion, from which kinematic distances may be derived.

Following \cite{Murphy13}, we used an iterative approach to find the kinematic distance (in 1~pc increments) of each star in Tables~\ref{tab:spectroscopy} and \ref{tab:previousmembers} which minimised the difference in space motion,
 \begin{equation}
\Delta v = [(U-U_{0})^2 + (V-V_{0})^2 + (W-W_{0})^2]^{\frac{1}{2}}
\end{equation}
where $(U,V,W)=f(\alpha,\delta, \mu, \textrm{RV}; d)$ and $(U_{0},V_{0},W_{0})$ is the mean velocity of the previous iteration \citep[initially from][]{Torres08}. Proper motions were sourced primarily from SPM4, with radial velocity measurements from WiFeS and the literature. As a check of the SPM4 proper motions we cross-checked its astrometry against UCAC4, Tycho-2 and PPMXL \citep{Roeser10}. In almost all cases the proper motions agreed at better than 2$\sigma$ and we adopted the SPM4 values. However, for CD$-$30 3394AB and candidates \#31 and \#61 we adopted proper motions from UCAC4, as these showed better agreement with Tycho-2 and PPMXL. We  similarly used the Tycho-2 proper motion for TYC 7066-1037-1 and the PPMXL values for the $\sim$1~arcmin double TYC 9300-0529-1 and TYC 9300-0891-1, as this was the only catalogue in which their proper motions agreed (and also agreed with Tycho-2 for TYC 9300-0891-1). \emph{Hipparcos} proper motions were also available for candidates \#25 and \#42, with the former having a parallax of $115\pm5$~pc, which we adopted in lieu of a kinematic distance.  

\begin{figure}
   \centering
   \includegraphics[width=0.9\linewidth]{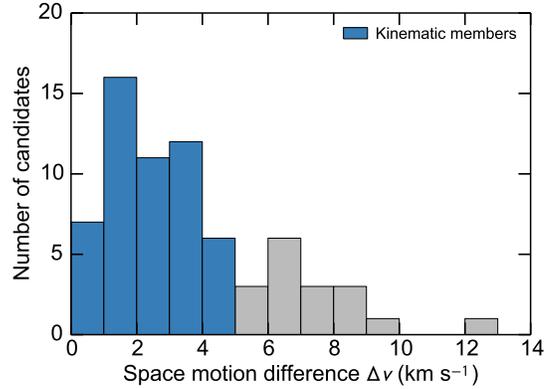} 
   \caption{Final distribution of $\Delta v$ values from the convergence analysis. We select as kinematic members those candidates with $\Delta v<5$~\kms. The tail of the distribution extends to 104~\kms\ for candidate \#44.}
   \label{fig:dkin}
\end{figure}

\begin{figure}
   \centering
   \includegraphics[width=0.9\linewidth]{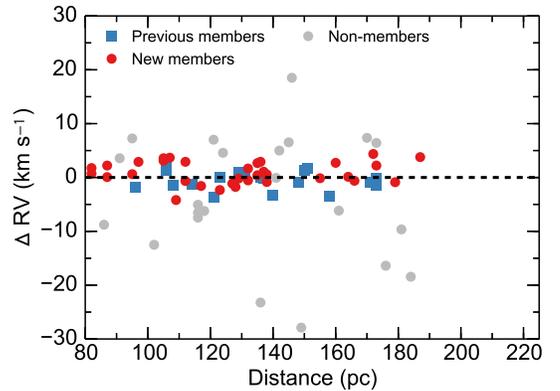}
   \caption{Difference between the observed radial velocity and that expected of an Octans member as a function of distance, for members proposed in the literature (blue squares) and our new members (red points). Grey points show those candidates with $\Delta v>5$~\kms\ not included in the final membership. The mean radial velocity difference is $+$0.4~\kms.}
   \label{fig:drv}
\end{figure}

Given the modest precision of the WiFeS velocities and the few \kms\  velocity dispersion typical of young associations, at each iteration we retained those candidates with $\Delta v<5$~\kms.  A new $(U_{0},V_{0},W_{0})$ was then generated and the process repeated until the velocity no longer changed. The resulting solution had 52 members and a mean space motion and standard deviation of
\begin{equation} \label{eqn:UVW}
(U,V,W)_{0} = (-13.2\pm1.5, -5.0\pm1.7, -11.0\pm1.5)~\kms 
\end{equation}
This is 1.9~\kms\ from the space motion reported by \citet{Torres08}  but agrees within the errors.  All 19 members proposed in the literature and 33/57 candidates were selected as kinematic members. The final distribution of $\Delta v$ values is plotted in Fig.~\ref{fig:dkin}, with the radial velocity residuals shown in Fig.~\ref{fig:drv}. We found no significant trends in these or the proper motion residuals with distance, sky position, heliocentric position or velocity, with the exception of an anti-correlation (Pearson's $r=-0.8$) between the $V$ velocity component and the radial velocity residual. 

\begin{figure}
   \centering
   \includegraphics[width=0.98\linewidth]{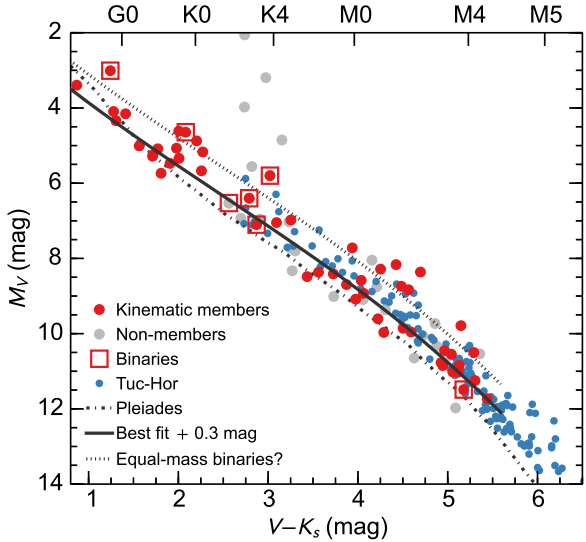} 
   \caption{Octans colour-magnitude diagram. The kinematic members (red, with non-members shaded grey) are fitted by a cubic (solid line) which has been shifted by $+$0.3~mag to better fit the main stellar locus, ostensibly tracing single stars or high mass ratio binaries. The dotted line (0.75~mag brighter)   shows the presumed position of equal-mass binary systems. Confirmed binaries are indicated. Distances to the Tuc-Hor members \citep[small blue points;][]{Kraus14} were derived from their SPM4 proper motions and the \citet{Torres08} space velocity.}
   \label{fig:cmd}
\end{figure}

Other reasonable initial space motions converged to similar final velocities ($\Delta v<0.5$~\kms) and membership lists. However, we found that the final velocity depended strongly on the convergence criteria adopted at each iteration. For example, a limit of $\Delta v<3$~\kms\ gave a final space motion 2.6~\kms\ from the first solution, inconsistent with its $U$ velocity at 1.3$\sigma$. This increased the kinematic distances by 10--30~pc, reducing the absolute magnitudes (Fig.~\ref{fig:cmd}) by 0.1--0.6~mag. Regardless of the criteria or initial velocity used, most of the solutions had $32\pm2$ ($51\pm2$) candidates within 3~(5)~\kms\ of the final velocity.  Trigonometric parallaxes of many Octans stars (e.g. from \emph{Gaia}) are necessary to firmly establish an accurate space motion for the group.  We adopt the $\Delta v<5$~\kms\ space motion of Eqn.~\ref{eqn:UVW} in the rest of the analysis, but caution that the true velocity may be several \kms\ from this value.  Space motions, kinematic distances and heliocentric positions for the stars selected as members are presented in Table~\ref{tab:members}. 

\subsection{Colour-magnitude diagram}

Using these kinematic distances, we plot in Fig.~\ref{fig:cmd} the Octans $M_{V}$ versus $V-K_{s}$ colour-magnitude diagram (CMD) and compare it to new K and M-type members of the $\sim$30~Myr Tucana-Horologium (Tuc-Hor) association from \cite{Kraus14}. Photometry for both groups was sourced from SPM4, except where its $V$ magnitudes were derived from photographic plates (SPM4 flag \texttt{iv=2}). For these stars we adopted the UCAC4/APASS-DR6 values. Photometry for all Octans candidates is listed in Tables~\ref{tab:spectroscopy} and \ref{tab:previousmembers}.  As expected from \S\ref{sec:cmdselect}, the new members generally lie above the Pleiades isochrone and fall on or slightly below the Tuc-Hor sequence down to the $V<18$~mag limit of our survey. While this suggests that Octans may have an age between Tuc-Hor and the Pleiades,  the uncertain kinematic distances and density of isochrones in this region of the diagram make it difficult to determine a precise age from photometry alone. 

Three confirmed visual (TYC 9300--0891--1, \#32) and spectroscopic (HD 155177) binaries are unresolved and lie above the majority of members in the CMD, close to the expected locus of equal-mass systems. This trend continues at later spectral types, with eight candidates at $V-K_{s}\gtrsim4$ following a similar sequence. Two of these stars (\#41, \#48) have broader CCFs (see \S\ref{sec:broadccf}) which may indicate spectroscopic binarity. The three other confirmed binary members lie on or below the single star sequence. Their proper motions (\#14, \#18) or radial velocities (\#16), and hence kinematic distances, may be affected by their close companions. It is also possible that some of the candidates in Figs.\,\ref{fig:dkin} and \ref{fig:drv} with $\Delta v>5$~\kms\ are spectroscopic binaries whose single-epoch velocities were non-systemic. Several of these stars lie close to the expected binary locus, though additional radial velocity observations are obviously required to confirm binarity.

\begin{figure*}
   \centering
   \includegraphics[width=0.48\linewidth]{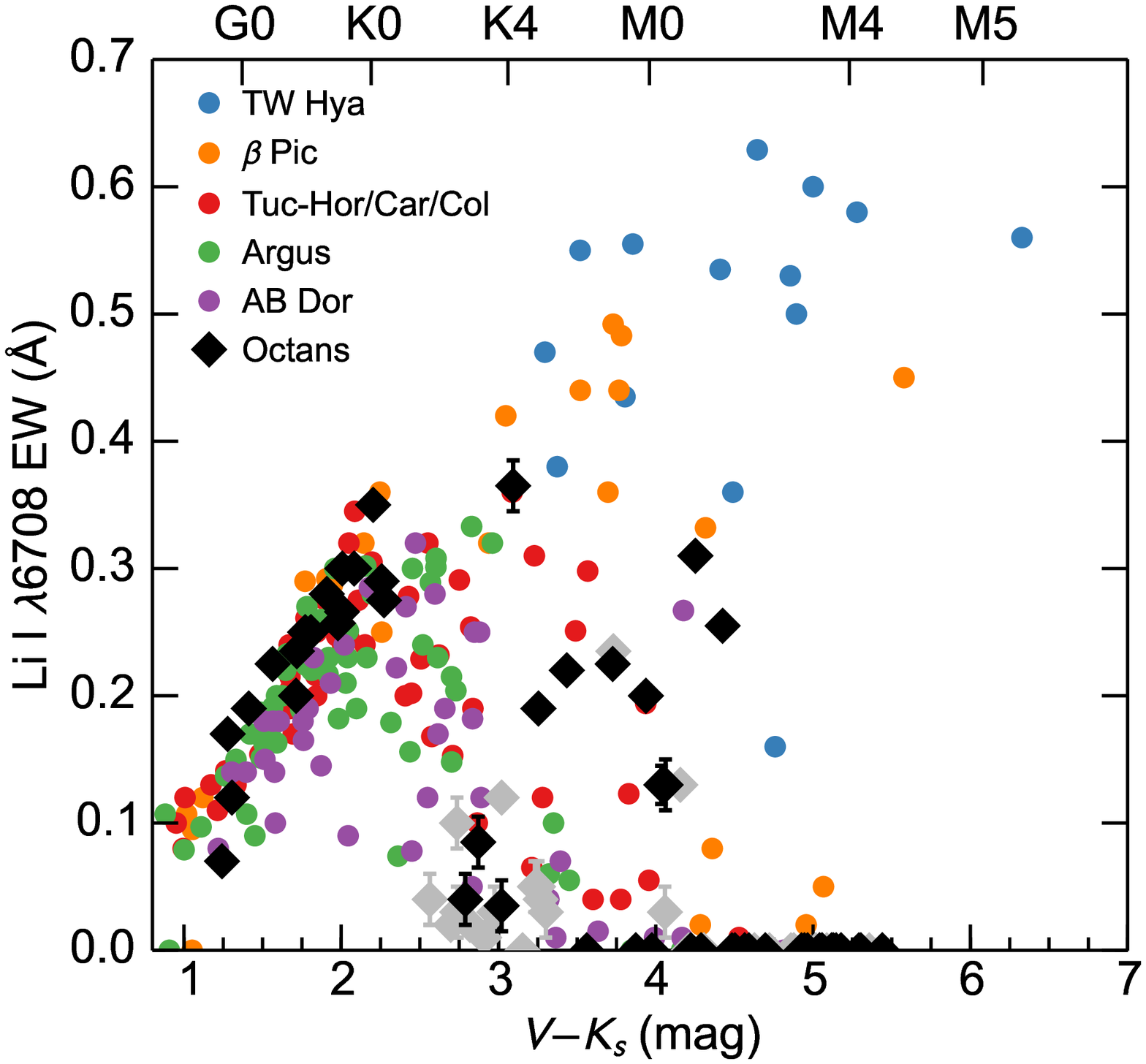}
   \includegraphics[width=0.48\linewidth]{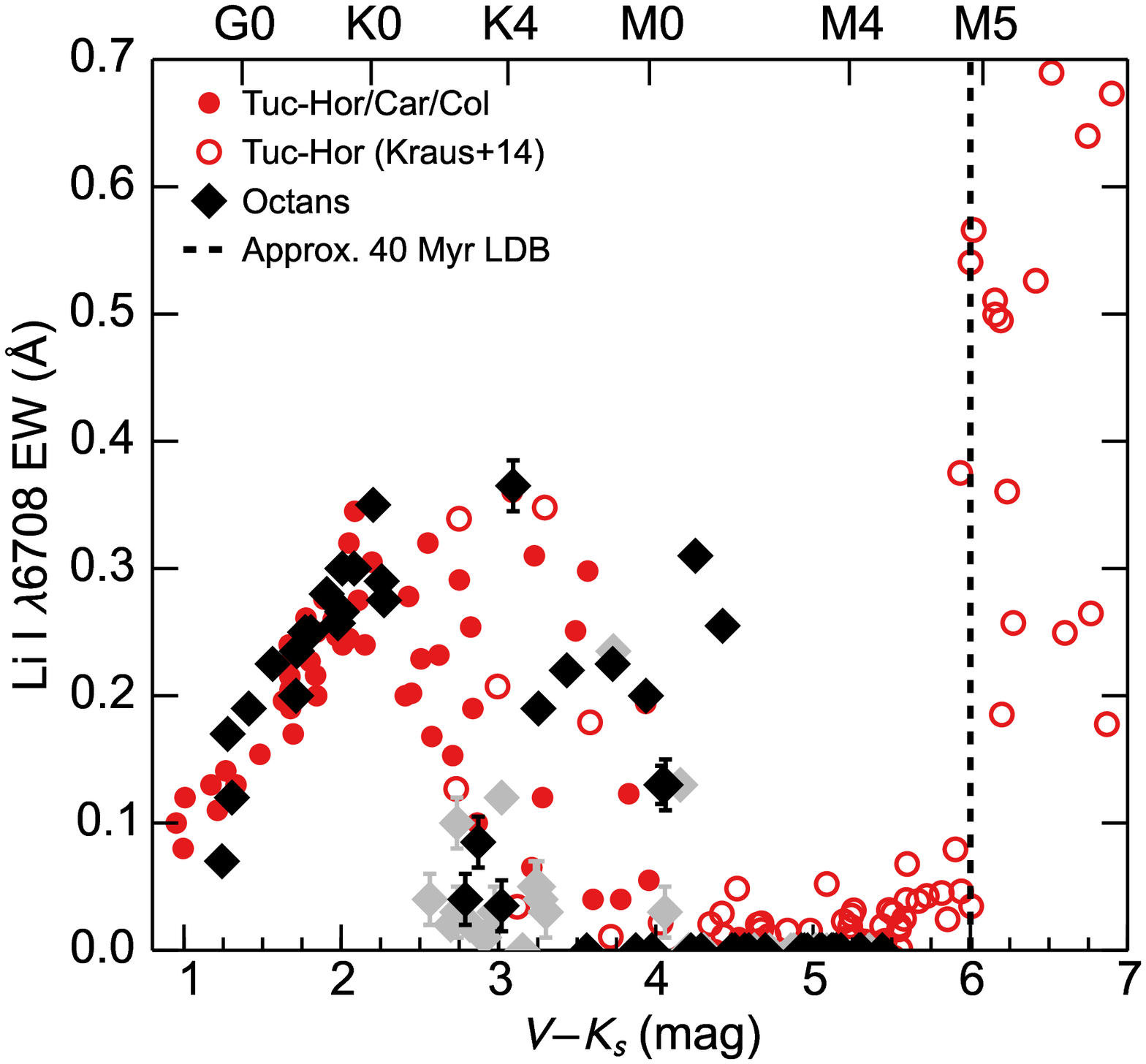} 
   \caption{\emph{Left:} Li\,\textsc{i} $\lambda$6708 equivalent widths (EWs) of Octans candidates (large diamonds) compared to young moving group members from \citet{da-Silva09}. Stars not selected as kinematic members are shaded grey. Several K and M-type Octans members have levels of lithium depletion between the $\beta$~Pictoris and Argus/ IC 2391 associations, similar to the Tucana-Horologium/Carina/Columba complex. The three kinematic members near $V-K_{s}=3$ are discussed in the text. \emph{Right:} Octans compared to Tuc-Hor/Car/Col members, with new Tuc-Hor members recently added by \citet{Kraus14} (open symbols). These stars span the $\sim$40~Myr lithium depletion boundary at a spectral type of M4.5--5 ($V-K_{s}\approx6$), just beyond the limit of our Octans survey. The EW upper limits on the WiFeS non-detections is 30--50~m\AA, similar to the scatter in the \citeauthor{Kraus14} measurements at $4<V-K_{s}<6$.}
   \label{fig:lithium}
\end{figure*}

\subsection{Lithium depletion}\label{sec:lithium}

The distribution of Li\,\textsc{i} $\lambda$6708 equivalent widths (EWs) is plotted in Fig.~\ref{fig:lithium}, where we compare Octans to other young moving groups studied by \citet{da-Silva09}. The G and early K-type members of Octans follow a similar pattern in this diagram to  other young groups, with minimal lithium depletion visible up to the $\sim$120~Myr age of the AB Doradus association \citep{Barenfeld13}. However, moving into the mid-K and M spectral types covered by our survey, there is evidence of significant depletion, bracketed between the $\sim$20~Myr $\beta$ Pictoris association \citep{Binks14}, and the older, more depleted Argus/IC 2391 \citep[30--50~Myr;][]{De-Silva13} and AB Doradus associations. The depletion pattern is similar to that observed in the $\sim$30~Myr Tuc-Hor association, which, together with the adjacent and co-eval Carina and Columba associations, loosely comprise the so-called Great Austral Young Association \citep[GAYA;][]{Torres01,Torres08}. 

In the right panel of Fig.~\ref{fig:lithium} we re-plot GAYA members from \citet{da-Silva09} with the new low-mass members of Tuc-Hor from Fig.~\ref{fig:cmd}. The similarities between Octans and Tuc-Hor are striking, with both groups showing a decline in EWs over the K spectral type range  to negligible lithium absorption by mid-M.  Across this temperature range there is considerable scatter in the EW measurements outside of instrumental errors. The mechanisms responsible for this star-to-star variation are not yet well understood, but may be related to early pre-main sequence mass-loss, rotation-induced mixing or the inhibition of convection by magnetic fields \citep{Soderblom10}. The sharp reemergence of the Li\,\textsc{i} $\lambda$6708 feature in Tuc-Hor stars at $V-K_{s}\gtrsim6$ is the so-called lithium depletion boundary (LDB) and is due to the extreme sensitivity of lithium burning to stellar mass in fully-convective pre-main sequence stars \citep{Basri96,Bildsten97}.  The temperature and luminosity \citep{Jeffries06} of the LDB changes with time as less massive stars begin lithium burning, allowing `semi-fundamental' model-insensitive age estimates for the handful of groups where the LDB has been detected \citep{Soderblom13}. Unfortunately our current spectroscopic survey does not span the Octans LDB, making it impossible to confirm that the association has an LDB age similar to the $40\pm2$~Myr estimated for Tuc-Hor by \citet{Kraus14}. 

Four candidates in Fig.~\ref{fig:lithium} with EW$_{\textrm{Li}}$$\geq$$100$~m\AA\ were not selected as kinematic members. Star \#31 ($\Delta v=18$~\kms) has an estimated distance of 186~pc and lies well above the expected binary sequence in Fig.~\ref{fig:cmd}. It is most likely an older pre-main sequence star with a proper motion similar to an Octans member at that distance. Candidates \#32A and \#36 have $\Delta v$ values of 6.9~\kms\ and 6.8~\kms, respectively, just outside the 5~\kms\ selection limit. Both systems lie near the equal-mass binary locus in Fig.~\ref{fig:cmd} and binarity may explain their larger $\Delta v$ values. Finally, candidate \#57 (EW$_{\textrm{Li}}=235$~m\AA) is only 5.4~\kms\ from the mean Octans space motion. The star lies just below the single star sequence in Fig.~\ref{fig:cmd}. Pending further investigation, we classify these three systems as possible members of Octans and list their particulars in Table~\ref{tab:possible}. Three kinematic members near $V-K_{s}=3$ have anomalously low EW values, more consistent with Argus or AB Doradus than other Octans members. All three systems (\#16, \#18B, \#32B) also show H$\alpha$ in absorption and are discussed in the next section.

\subsection{Activity and X-ray emission}\label{sec:activity}

\begin{figure}
   \centering
   \includegraphics[width=0.95\linewidth]{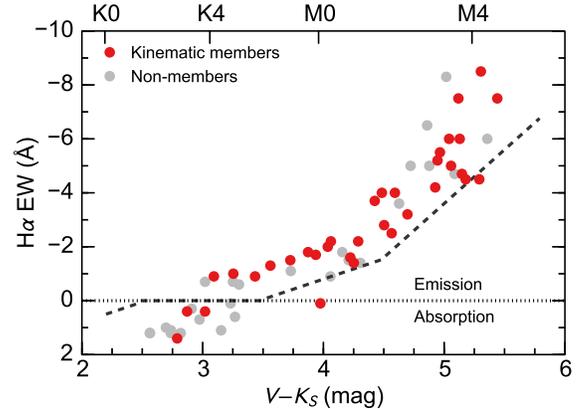} 
   \caption{WiFeS H$\alpha$ EWs for Octans kinematic members (red) and non-members (grey). Following \citet{Kraus14}, the dashed line shows the lower envelope of EWs observed in the 40--50~Myr open clusters IC 2391 and IC 2602 \citep{Stauffer97}, converted from $V-I_{C}$ to $V-K_{s}$. The four members with $\textrm{EW}>0$ are discussed in the text.}
   \label{fig:halpha}
\end{figure}

All of the Octans candidates were selected as NUV-bright, presumably active stars in \S\ref{sec:nuv}. In the wavelength range of our WiFeS spectra, the best activity indicator is the Balmer H$\alpha$ line, which should be in emission  across the spectral types and ages considered in this survey \citep{Zuckerman04a}. However, because a non-negligible fraction of nearby older K and M-type field stars have been found to exhibit elevated activity levels \citep[up to 30--40 per cent for M4 stars at $|Z|<100$~pc;][]{West11}, the presence of H$\alpha$ emission is a necessary but insufficient indicator of youth. We plot in Fig.~\ref{fig:halpha} the H$\alpha$ EWs of the WiFeS candidates, with the lower activity bound observed in the 40--50~Myr open clusters IC 2391 and IC 2602  \citep{Stauffer97}. As expected of young, active objects, the majority of Octans members lie on or above the cluster locus. However, four kinematic members show H$\alpha$ in absorption. We immediately classify star \#37 (M0; EW$_{\textrm{H}\alpha}=0.1$~\AA) as an older field star with similar kinematics to Octans ($\Delta v=4.7$~\kms). Despite marginal lithium detections in both components of \#18AB, their strong H$\alpha$ absorption ($1.3/1.4$~\AA) rules out a young age. The system is probably of Pleiades age or older.  The two K-type members just below the $\textrm{EW}_{\textrm{H}\alpha}=0$ threshold are the SB1 \#16  and \#32B. Both \#16 and \#32AB are lithium rich, and while the former showed H$\alpha$ core emission in its 2014 Jan 17 spectrum, \#32A has clear H$\alpha$ emission ($\textrm{EW}_{\textrm{H}\alpha}=-0.7$~\AA). Given these mixed characteristics, we retain them as possible members of Octans in Table~\ref{tab:possible}.

\begin{figure}
   \centering
   \includegraphics[width=0.9\linewidth]{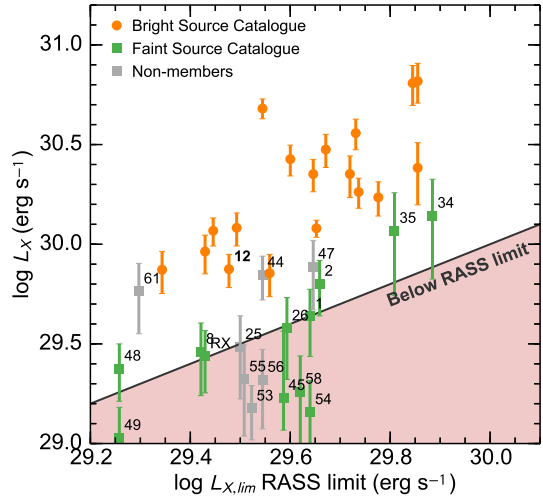} 
   \caption{\emph{ROSAT} All Sky Survey (RASS) 0.1--2.4 keV X-ray luminosities of proposed Octans members (orange, green) and kinematic non-members (grey).  Approximate uncertainties were calculated from the uncertainties on the X-ray count rate and HR1 hardness ratio. The abscissa gives the approximate RASS luminosity limit at the kinematic distance of each star, corresponding to a limiting flux of $2\times10^{-13}$~erg cm$^{-2}$~s$^{-1}$ \citep{Schmitt95}. Assuming their X-ray fluxes and distances are correct (see text), the candidates in the shaded region of the diagram are fainter than this limit and should not have been detected by RASS.}
   \label{fig:rosat}
\end{figure}

Coronal X-ray emission is another common manifestation of stellar activity, and hence youth \citep{Preibisch05}. Indeed, all of the solar-type Octans members discovered by SACY \citep{Torres08,Elliott14} were selected as optical counterparts to \emph{ROSAT} X-ray sources. However, given the larger mean distance of Octans we do not expect many K and M-type members to have been observed by the satellite (see discussion in \S\ref{sec:introduction}). To check for unusually bright or nearby X-ray sources, we cross-matched Tables~\ref{tab:spectroscopy} and \ref{tab:previousmembers} against the \emph{ROSAT} All Sky Survey (RASS) Bright and Faint Source Catalogues \citep{Voges99,Voges00}\footnote{\texttt{ivo://nasa.heasarc/rassbsc}(\texttt{rassfsc})} with a search radius of 2~arcmin. The results are shown in Fig.~\ref{fig:rosat}, where we have converted the observed count rates and hardness ratios (HR1) to X-ray luminosities using the kinematic distance of each candidate and the conversion factor of \citet{Fleming95b}. In addition to the 17 SACY members and \rxjstar, the lithium-rich kinematic member \#12 was matched to the BSC source 1RXS J062045.9--362003 and 18 other candidates were matched to FSC sources. The 12 candidates (including 8 new kinematic members) above the RASS detection limit are listed in Table~\ref{tab:rosat}, with their estimated 0.1--2.4~keV luminosities and $\log L_{X}/L_{\textrm{bol}}$ flux ratios. With the exception of the non-member \#25 (matched to the same source as \#26), these values are consistent with the saturated X-ray emission commonly seen in young, low-mass stars \citep{Zuckerman04a}.  The `super-saturated' fractional X-ray emission observed in kinematic member \#2 ($\log L_{X}/L_{\textrm{bol}}=-2.2$) is unsurprising given the M3.5 star was detected well above the RASS limit at a distance of 143~pc. The coincidence of its optical and X-ray positions is confirmed from a high positional accuracy (4.3~arcsec) serendipitous observation by the \emph{Swift} satellite \citep{Evans12}. Seven stars matched to \emph{ROSAT} sources lie below the RASS limit in the shaded region of the diagram, including the lithium-rich member \#45, which was only detected in the soft X-ray band ($\textrm{HR1}=-1$). Assuming its true HR1 is $\sim$0 \citep[typical of young moving group members;][]{Kastner03}, we find an X-ray luminosity on the detection limit and $\log L_{X}/L_{\textrm{bol}}=-3.27$. The remaining candidates have extreme HR2 values (\#49, \#58; which also affect HR1), may be confused with other nearby sources (\#54), or are kinematic non-members with inappropriate distances.

\subsection{Infrared excesses}\label{sec:SED}

In contrast to the much higher \citep[20--40~per cent;][]{Wyatt08,Zuckerman11} debris disc fractions observed in young solar-type stars, mid-infrared excesses due to debris discs are rare around late-K and M-dwarfs older than $\sim$20~Myr \citep[see discussion and references in][]{Deacon13}. The SED analysis in \S\ref{sec:sptypes} identified three candidates (\#1, \#11, \#12) with weak (2--3$\sigma$) excess emission above photospheric levels in the \emph{WISE} $W4$ (22~\micron) passband. A similar analysis for the 19 members proposed in the literature  found four excess sources, including the known debris disc host HD 36968. However, after inspection of the individual and co-added $W4$ images and considering the small excesses and low signal-to-noise ($<$5) of the \emph{WISE} detections at these magnitudes ($W4>8$~mag), only HD 36968 (spectral type F2) is likely to have a real excess at 22~\micron. Its \emph{WISE} flux is consistent with the \emph{Spitzer Space Telescope} 24~\micron\  MIPS measurement and IRS spectrum presented by \citet{Moor11}. The other apparent excess sources were only marginally detected in the co-added images and their $W4$ photometry cannot be trusted. We calculate an upper limit disc fraction for Octans of 3$_{-1}^{+7}$~per cent (1/30), where the 68 per cent confidence interval was calculated using the method of \citet{Cameron11} and we considered only those 30 stars in Table~\ref{tab:members} with $W4$ measurements. Our findings are consistent with the conclusions of \citet{Deacon13} and the results of \citet{Kraus14}, who found no strong excess sources in their sample of K3--M6 Tuc-Hor members at a similar age to Octans.

\section{Discussion}\label{sec:discussion}

\subsection{Final membership solution}\label{sec:finaluvw}

The final Octans membership solution is listed in Table~\ref{tab:members}, where we have removed the possible members \#16 and \#32AB as well as the non-members \#18AB and \#37 (\S\ref{sec:activity}). After re-running the convergence analysis, 48 members were retained and the mean space motion converged to
\begin{equation} \label{eqn:newUVW}
(U,V,W)_{0} = (-13.7\pm1.2, -4.8\pm1.7, -10.9\pm1.5)~\kms 
\end{equation}
This is only 0.5~\kms\ from the velocity in Eqn.~\ref{eqn:UVW} and increased the kinematic distances of the members by $\lesssim$6~pc. During the re-analysis the marginal kinematic member \#30 ($\Delta v=4.6$~\kms) fell just outside of the 5~\kms\ limit and was replaced by star \#55, which was not in the initial solution but finished with $\Delta v=4.9$~\kms. All other members were retained. As emphasized in \S\ref{sec:kinematicmembers}, this is our best estimate of the Octans velocity given the available observations. However, the absence of any trigonometric parallaxes means the final velocity may change by several \kms\ once these are available. 

\subsection{Age of Octans}

The oft-cited age of Octans is $\sim$20~Myr, based loosely on the 15 solar type stars presented in \citet{Torres08}. A lithium age is most useful in this instance as it is distance-independent, and hence not influenced by changes in the group space motion which affect the kinematic distances. We find that the new low-mass members identified in this work show  a lithium depletion pattern similar to the Tuc-Hor association (Fig.~\ref{fig:lithium}), which has recently had a lithium depletion boundary (LDB) age of $40\pm2$~Myr estimated by \citet{Kraus14}. Octans is clearly older than the $\beta$ Pictoris association \citep[LDB age $21\pm4$~Myr;][]{Binks14} but younger than the Argus/IC 2391 association \citep[LDB age $50\pm5$~Myr;][]{Barrado-y-Navascues04}.  While all of these groups have differing age estimates from other methods (for example low-mass isochrone or main sequence turn-off fitting), their relative ages and ranking are sound, i.e. $\tau_{\beta \textrm{ Pic}}<\tau_{\textrm{Octans}}\approx\tau_{\textrm{Tuc-Hor}}<\tau_{\textrm{IC 2391}}$. Distance uncertainties notwithstanding, the Octans colour-magnitude diagram (Fig.\,\ref{fig:cmd}) is also consistent with the above age range, as are the more qualitative age indicators (possible fast rotators, H$\alpha$ and X-ray activity, lack of protoplanetary discs). For the remaining discussion we adopt an age of 30--40~Myr, in agreement with the LDB age estimate for Tuc-Hor but also consistent with the younger (20--30~Myr) isochronal age of that group. The discovery of lithium-rich Octans members on the faint side of its LDB (spectral type M4/5) would immediately allow a more robust age estimate (see \S\ref{sec:lithium}). Recent work has demonstrated that the well-established discrepancy between the LDB and isochronal ages of young groups \citep[e.g.][]{Pecaut12,Yee10,Song02a} may be due to an incomplete treatment of magnetic fields in pre-main sequence models \citep{Malo14}. 

\subsection{Structure and relationship to Octans-Near}\label{sec:structure}

\begin{figure*}
   \centering
   \includegraphics[width=0.98\linewidth]{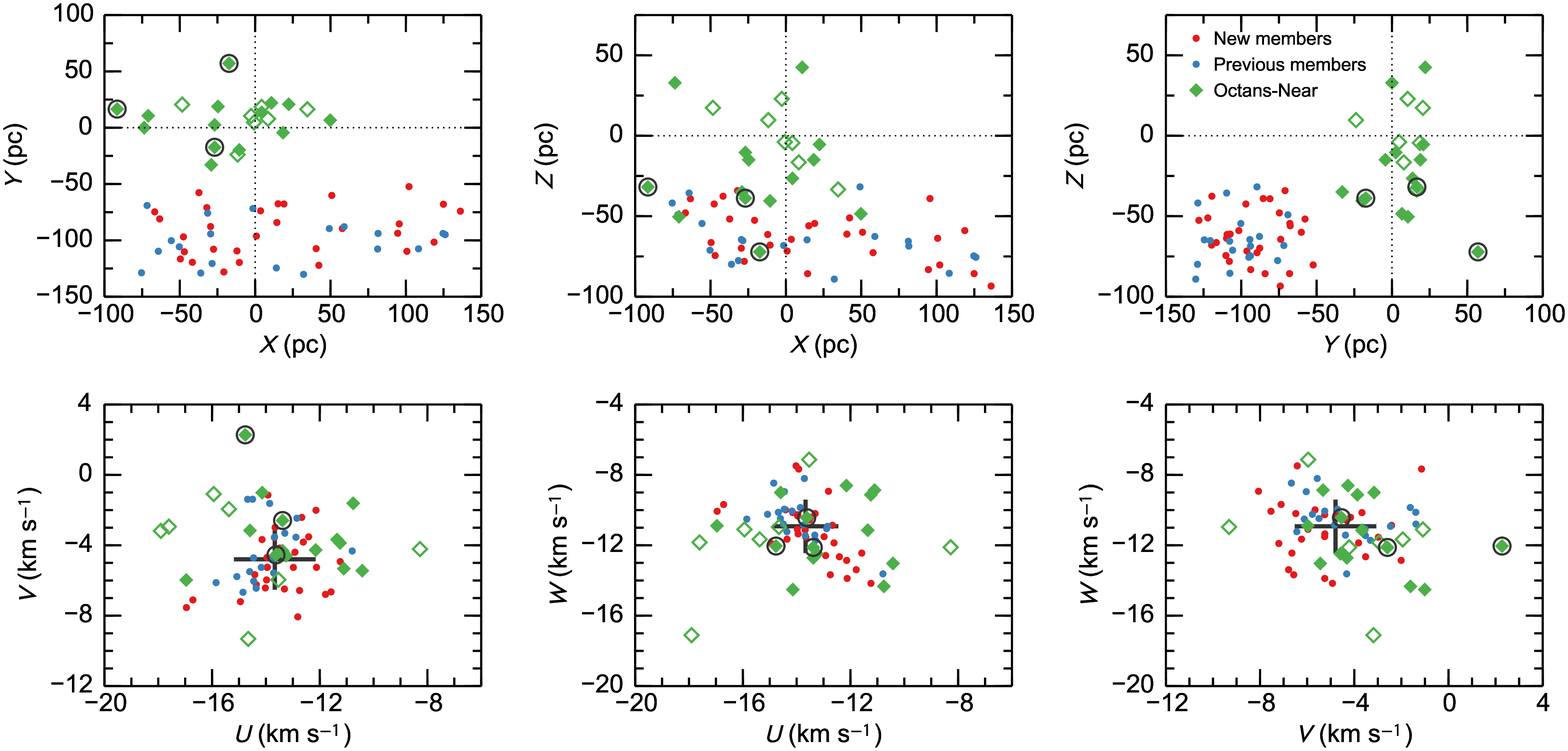} 
   \caption{Heliocentric  positions (top row) and velocities (bottom row) for new Octans members identified in this work (red points), literature members (blue points) and Octans-Near stars from \protect\cite{Zuckerman13} (green diamonds; open symbols are possible members). Octans data were taken from the kinematic solution in Table~\ref{tab:members}, while positions and velocities for the Octans-Near stars were calculated from their \emph{Hipparcos} astrometry and literature radial velocities  \protect\citep[tabled in][]{Zuckerman13}. The three spectroscopically young ($\sim$30~Myr) Octans-Near members are marked by circles. The position of the Sun is given by the dotted lines, while the error bars show the mean Octans space motion and its standard deviation (Eqn.~\ref{eqn:newUVW}).}
   \label{fig:xyzuvw}
\end{figure*}

The heliocentric positions of Octans members from Table~\ref{tab:members} are plotted in Fig.~\ref{fig:xyzuvw}. Their mean distance is $132\pm25$~pc, with a range of 85~pc (star \#11) to 181~pc (\#34). \citet{Torres08} already noted that the association is unusually extended in the $X$ dimension, and our new members follow an identical distribution. Unlike the Tuc-Hor, AB Dor or $\epsilon$ Cha associations, no central core or nucleus is readily apparent. Octans spans over 200~pc in $X$, but is similar to other nearby associations in $\Delta Y$ (80~pc) and $\Delta Z$ (60~pc). 

What could cause this sparse, cigar-like structure aligned towards the Galactic centre?  Octans is too young to have been significantly modified by the Galactic potential, which would normally manifest as an elongation in $Y$ due to differential rotation, accentuated by Octans' large size (0.2~kpc) in $X$. Hence its present-day structure must trace the dynamical state of its birth cloud and any local processes that have acted on the stars since. Neglecting individual observational errors and assuming that the $\sigma_{U}=1.2$~\kms\ in Eqn.~\ref{eqn:newUVW} is wholly due to the intrinsic velocity dispersion of the association \citep[see discussion in][]{Mamajek14}, it is impossible to explain the observed size of Octans assuming a $\sim$30~Myr age. Nor can linear expansion account for its present-day configuration. We find only a weak correlation (Pearson's $r=0.44$) between $U$ and $X$, with a gradient of 8.4$\times$10$^{-3}$~\kms\ pc$^{-1}$. This corresponds to an expansion age of $\sim$120~Myr, similar to the $\sim$175~Myr crossing time ($\Delta X/\sigma_{U}$). To reach the size observed today, star formation in Octans must have proceeded in a distributed, dynamically quiet manner over many tens of parsec, similar to an OB or sparse T association. An analogy can be made to the $\sim$30~Myr-old GAYA complex \citep[Great Austral Young Association;][]{Torres01,Torres08}, which loosely incorporates several hundred approximately co-eval members of the Tuc-Hor, Columba and Carina associations with other young stars over a similar volume. However, unlike GAYA the current membership of Octans is much sparser and lacks its kinematic and spatial substructure.

Because our survey was restricted to the volume occupied by existing members, it is possible that Octans extends further than the stars in Fig.~\ref{fig:xyzuvw}.  Recent work  hints that this may be true. While investigating the nearby young star HIP 17338\footnote{\citet{Viana-Almeida09} classify this star as a member of Octans, but it is not in the memberships of \citet{Torres08} or \citet{da-Silva09}.}, \citet{Zuckerman13} noted its space motion matched that of Octans but at a much smaller distance ($d_{\textrm{Hip}}=50$~pc). They subsequently identified 14 probable and seven possible systems with similar space motions and \emph{Hipparcos} parallaxes within 100~pc, calling this new group `Octans-Near'.  These stars are also plotted in Fig.~\ref{fig:xyzuvw}. While there is some overlap in $Z$ (Octans-Near spans the Galactic plane), the two groups are distinct in $X$ and $Y$. Given the $d<100$~pc limit adopted by \citeauthor{Zuckerman13} and the constrained region of our survey, this separation may be somewhat artificial and the two groups could occupy a much larger volume. In this case the many UV-bright kinematic candidates from \S\ref{sec:nuv} will be useful for identifying new members.  The mean space motion of the 14 probable Octans-Near members tabled by \citet{Zuckerman13} is $(U,V,W) = (-13.0\pm1.9, -3.5\pm2.2, -11.2\pm2.0)$~\kms. Including these stars and the seven possible Octans-Near members in the convergence analysis yields the same mean space motion as Eqn.~\ref{eqn:newUVW} ($\Delta v<0.2$~\kms) with a slightly larger error in $U$ (1.5~\kms) and adds 16 new members to a much larger, but kinematically-coherent Octans `complex'.  However, only three Octans-Near stars (HIP 813, HIP 17338 and HIP 19496; marked in Fig.~\ref{fig:xyzuvw}) have ages estimated by \citeauthor{Zuckerman13} from lithium and activity as young as the $\sim$30~Myr age of Octans. The other proposed members are approximately Pleiades age or older. 

Despite the kinematic similarities, the large age range in Octans-Near and the groups' grossly different spatial (and sky) positions mean any relationship between Octans and Octans-Near remains unclear. Given the multitude of pre-main sequence stars of various ages with velocities around Eqn.~\ref{eqn:newUVW}, it may be more useful to think of the area as a recent star-forming complex, in which some stars are as young as 30~Myr, while the majority are older \citep{Zuckerman13}. In this scenario the co-evality of Octans as it is defined here is not guaranteed and many of the late-type members in Table~\ref{tab:members} without lithium detections may be older than the nominal 30--40~Myr. Furthermore, any kinematic substructure across the association would render our simple convergence model and its distances incorrect. Accurate trigonometric parallaxes and radial velocities for many late-type members from \emph{Gaia} will be vital to mapping the full structure of the complex and resolving the degeneracies between space motion, kinematic distance and age. 

\section{Conclusions}

To better determine its age and bulk properties, we have undertaken the first systematic survey for K and M-type members of the under-studied Octans association. The main results of this work are:

(1) From the spectroscopic follow-up (spectral types, radial velocities, Li\,\textsc{i}, H$\alpha$) of 57 photometric and proper motion candidates, we identified 29 new K5--M4 members of Octans (Table~\ref{tab:members}), doubling the existing solar-type membership. Four more systems are possible Octans members requiring further investigation. Several new members  are spectroscopic binaries or show broader spectral features suggestive of rapid rotation. 

(2) Combining these stars with other members proposed in the literature, we undertook a convergence analysis to refine the space motion and kinematic distances of the group. Our Octans solution has 48 members at distances of 85--181~pc and a mean velocity of $(U,V,W)= (-13.7\pm1.2, -4.8\pm1.7, -10.9\pm1.5)$~\kms.  This velocity should be adopted over previous estimates as it is derived from more members over a wider range of spectral types.

(3) Lithium depletion in the new K and M-type members is similar to that seen in the 30--40~Myr-old Tucana-Horologium association. Octans is clearly older than the $\sim$20~Myr $\beta$ Pictoris association and is younger than the Argus/IC 2391 (40--50~Myr) and AB Doradus ($\sim$120~Myr) associations. 

(4) Octans is very elongated in $X$ (200~pc) and much larger than other nearby groups of similar age. A dozen or so members of the foreground `Octans-Near' association \citep{Zuckerman13} with \emph{Hipparcos} parallaxes have similar kinematics to Octans, but only three stars are young enough to conceivably be members of Octans as it is defined here. The two associations may form part of a larger complex which has seen several episodes of star formation over the past 100~Myr. Accurate parallaxes and velocities for many more members are needed to test this hypothesis.

\section*{Acknowledgments}

We thank Mike Bessell (ANU) for observing several candidates in 2014 April and Andy Casey (Cambridge) for providing the Python cross-correlation code, as well as the referee for their prompt and thorough review.  We also thank the ANU TAC for their generous allocation of telescope time.  SJM thanks Margaret Wilson for useful discussions on many aspects of this work and acknowledges support from a Gliese Fellowship at the University of Heidelberg. This research has made extensive use of the VizieR and SIMBAD services provided by CDS, the TOPCAT software package \citep{Taylor05} and Astropy, a community-developed core Python package for Astronomy \citep{Astropy-Collaboration13}. The TAP service queried in this paper was constructed as part of the activities of the German Astrophysical Virtual Observatory and VOSA is developed under the Spanish Virtual Observatory project supported from the Spanish MICINN through grant AyA2008-02156.  This publication makes use of data products from \emph{GALEX}, operated for NASA by the California Institute of Technology; the \emph{Wide-field Infrared Survey Explorer}, a joint project of the University of California, Los Angeles, and the Jet Propulsion Laboratory/California Institute of Technology, funded by NASA; and 2MASS, a joint project of the University of Massachusetts and the Infrared Processing and Analysis Center/California Institute of Technology, funded by NASA and the National Science Foundation.

\footnotesize{
\bibliographystyle{mn2e}

}

\input{table_spectroscopy.tex}
\input{table_previousmembers.tex}
\input{table_members.tex}
\input{table_possible.tex}
\input{table_rosat.tex}

\end{document}

%% file: ads_defns.tex
\newcommand{\aj}{AJ}                   
\newcommand{\araa}{ARA\&A}             
\newcommand{\apj}{ApJ}                 
\newcommand{\apjl}{ApJ}                

\newcommand{\apjs}{ApJS}               
     
\newcommand{\apss}{Ap\&SS}             
\newcommand{\aap}{A\&A}                
  
\newcommand{\aaps}{A\&AS}              
\newcommand{\mnras}{MNRAS}             
\newcommand{\iaucirc}{IAU~Circ.}       

%% file: table_spectroscopy.tex
\begin{table*}
\caption{Results of the ANU 2.3-m/WiFeS spectroscopy for the 57 Octans candidates lying above the Pleiades isochrone in Fig.\,\ref{fig:cmdselect}. Candidates 5, 6, 7 and 10 were not observed. $K_{s}$ magnitudes for all candidates are taken from 2MASS, with $V$ magnitudes and proper motions from SPM4, except for candidates 3, 4, 12, 43AB, 47 ($V$ from UCAC4), 61 (proper motion and $V$ from UCAC4), 31 (proper motion from UCAC4), 25 and 42 (proper motions from \emph{Hipparcos}). Candidates 18, 32 and 43 were marginally resolved as close doubles in the WiFeS field of view, but are unresolved in all-sky surveys. Spectroscopic binaries and the nine candidates with broader cross-correlation functions (CCF) are indicated in the final column. \label{tab:spectroscopy}}
\begin{tabular}{lcccccccccc}
\hline
ID & 2MASS designation & $V$ & $K_{s}$ & $\mu_{\alpha}\cos\delta$ & $\mu_{\delta}$ & Spectral & RV & Li EW & H$\alpha$ EW & Remarks \\
 & (J2000) & (mag) & (mag) & (mas yr$^{-1}$) & (mas yr$^{-1}$) & Type & (km s$^{-1}$) & (m\AA) & (\AA) &  \\
\hline
1 & 05321635--4132446 & 14.58 & 10.52 & $-$6.8 & 15.6 & M1 & 14.5 & 130 $\pm$ 20 & $-$2.2 & -- \\
2 & 05541279--4044059 & 16.55 & 11.61 & $-$10.8 & 13.8 & M3.5 & 12.9 & $<$50 & $-$5.2 & -- \\
3 & 04591776--3443443 & 13.22 & 9.95 & $-$11.8 & 14.2 & K6 & 22.9 & 40 $\pm$ 10 & 0.6 & -- \\
4 & 04593927--3426546 & 14.54 & 10.25 & $-$8.3 & 20.6 & M1.5 & 16.4 & $<$50 & $-$2.2 & -- \\
8 & 05410719--2409179 & 13.59 & 10.16 & $-$8.1 & 12.5 & K6 & 19.6 & 220 $\pm$ 10 & $-$0.9 & -- \\
9 & 06555474--4046498 & 15.41 & 10.91 & $-$17.8 & 12.9 & M2 & 11.6 & $<$40 & $-$2.8 & -- \\
11 & 06124600--3836083 & 16.30 & 10.86 & $-$21.8 & 20.7 & M4 & 15.3 & $<$50 & $-$7.5 & Broad CCF \\
12 & 06204718--3619482 & 12.29 & 9.20 & $-$16.8 & 14.0 & K5 & 16.5 & 365 $\pm$ 20 & $-$0.9 & Broad CCF \\
13 & 06055373--2541242 & 15.95 & 10.82 & $-$12.1 & 13.4 & M3.5 & 18.3 & $<$40 & $-$6.0 & -- \\
14 & 16405331--7234127 & 17.09 & 11.91 & $-$2.2 & $-$27.5 & M4 & $-$1.4 & $<$40 & $-$4.5 & SB2 \\
15 & 17325153--7336111 & 13.08 & 9.83 & 6.4 & $-$21.6 & K5 & $-$3.1 & 190 $\pm$ 10 & $-$1.0 & -- \\
16 & 14041946--7544411 & 12.29 & 9.42 & $-$20.1 & $-$28.2 & K3 & $-$5.6 & 85 $\pm$ 20 & 0.4 & SB1/2 \\
17 & 12234891--8115263 & 10.66 & 7.51 & $-$20.3 & $-$16.1 & K5 & 7.4 & $<$30 & 1.1 & -- \\
18A & 16214871--7726191 & 12.76 & 9.97 & 2.9 & $-$19.8 & K3 & 46.2 & 40 $\pm$ 20 & 1.3 & 4$^{\prime\prime}$ dbl. \\
18B & -- & -- & -- & -- & -- & K3 & 2.2 & 40 $\pm$ 20 & 1.4 & -- \\
19 & 17230629--7718207 & 16.02 & 11.16 & 7.0 & $-$19.6 & M3 & $-$11.1 & $<$40 & $-$6.5 & Broad CCF \\
20 & 17492993--7655068 & 13.16 & 10.46 & 6.8 & $-$20.2 & K3 & $-$17.8 & 20 $\pm$ 10 & 1.0 & -- \\
21 & 17414150--7826161 & 11.94 & 9.12 & 7.5 & $-$18.5 & K4 & 50.0 & 20 $\pm$ 10 & 1.2 & -- \\
22 & 17483458--7918534 & 16.33 & 11.31 & 16.4 & $-$19.6 & M3.5 & $-$0.8 & $<$50 & $-$8.3 & -- \\
23 & 17214553--8026396 & 16.57 & 11.53 & 7.1 & $-$22.5 & M3.5 & 2.2 & $<$40 & $-$6.0 & -- \\
24 & 08145586--7613014 & 14.08 & 10.35 & $-$24.7 & 10.3 & K7 & 7.3 & 225 $\pm$ 10 & $-$1.5 & -- \\
25 & 06205379--7903599 & 7.36 & 4.62 & $-$26.93 & 33.51 & K4 & 90.1 & 30 $\pm$ 20 & 1.1 & -- \\
26 & 06205027--7905195 & 14.37 & 9.80 & $-$14.9 & 24.1 & M1.5 & 3.4 & $<$40 & $-$2.5 & -- \\
27 & 06002910--5531015 & 15.90 & 10.94 & $-$15.0 & 19.1 & M3 & 8.6 & $<$40 & $-$5.5 & Broad CCF \\
28 & 06434892--5130461 & 15.29 & 11.07 & $-$15.9 & 15.5 & M1.5 & 11.7 & $<$40 & $-$1.6 & -- \\
29 & 18541290--6539499 & 14.52 & 10.31 & 12.1 & $-$22.5 & M1.5 & 1.5 & $<$40 & $-$1.5 & -- \\
30 & 18321860--7234381 & 14.54 & 9.84 & 13.6 & $-$16.9 & M2.5 & 2.2 & $<$40 & $-$3.2 & -- \\
31 & 19003118--7326569 & 10.30 & 7.56 & 9.2 & $-$17.6 & K3 & $-$20.0 & 100 $\pm$ 20 & 1.2 & -- \\
32A & 19223387--6722019 & 11.99 & 8.97 & 16.5 & $-$13.9 & K4 & 3.9 & 120 $\pm$ 10 & $-$0.7 & 3$^{\prime\prime}$ dbl. \\
32B & -- & -- & -- & -- & -- & K4 & $-$0.3 & 35 $\pm$ 20 & 0.4 & -- \\
33 & 19543436--6617340 & 15.22 & 11.16 & 19.0 & $-$11.8 & M1 & 35.0 & 30 $\pm$ 20 & $-$0.9 & -- \\
34 & 19545407--6450343 & 14.55 & 10.30 & 18.2 & $-$10.1 & M1 & $-$3.1 & 310 $\pm$ 10 & $-$1.4 & -- \\
35 & 19545336--6450090 & 14.24 & 9.81 & 17.2 & $-$14.8 & M1.5 & $-$2.1 & 255 $\pm$ 10 & $-$3.7 & -- \\
36 & 19353962--7047361 & 14.08 & 9.92 & 20.1 & $-$10.8 & M1 & $-$7.7 & 130 $\pm$ 10 & $-$1.8 & -- \\
37 & 19563472--7027160 & 14.23 & 10.26 & 30.9 & $-$16.6 & M0 & 2.5 & $<$40 & 0.1 & -- \\
38 & 19590891--8311502 & 16.69 & 11.33 & 20.0 & $-$8.3 & M4 & 8.5 & $<$40 & $-$6.0 & -- \\
39 & 20410187--7813502 & 15.74 & 10.59 & 23.3 & $-$5.3 & M4 & 0.7 & $<$40 & $-$4.7 & -- \\
40 & 22540018--7807039 & 13.63 & 9.76 & 36.6 & 10.1 & M0 & 5.8 & $<$30 & $-$1.8 & -- \\
41 & 20293579--6317079 & 14.44 & 9.96 & 24.3 & $-$12.4 & M2 & $-$0.9 & $<$30 & $-$4.0 & Broad CCF \\
42 & 21181373--7100299 & 9.06 & 6.09 & 24.42 & $-$6.92 & K5 & $-$27.1 & 30 $\pm$ 20 & 0.7 & -- \\
43A & 23465209--7359482 & 11.85 & 9.29 & 27.7 & 14.7 & K4 & $-$2.4 & 40 $\pm$ 20 & 1.2 & 4$^{\prime\prime}$ dbl. \\
43B & -- & -- & -- & -- & -- & K4 & $-$3.3 & 40 $\pm$ 20 & 1.2 & -- \\
44 & 04580405--8429420 & 12.44 & 9.21 & $-$8.9 & 29.0 & K5 & 107.6 & 50 $\pm$ 20 & 0.1 & -- \\
45 & 00530603--8250259 & 13.24 & 9.30 & 21.8 & 19.3 & M0 & 2.6 & 200 $\pm$ 10 & $-$1.7 & -- \\
46 & 05153537--7605430 & 16.47 & 11.84 & $-$1.5 & 23.6 & M3.5 & 24.9 & $<$40 & $-$3.6 & -- \\
47 & 04473187--7412323 & 13.40 & 10.14 & $-$2.4 & 25.3 & K5 & $-$16.2 & 50 $\pm$ 20 & $-$0.7 & -- \\
48 & 04085395--7203535 & 15.20 & 9.91 & $-$3.9 & 38.7 & M3.5 & 9.9 & $<$30 & $-$4.5 & Broad CCF \\
49 & 04105436--6924209 & 15.47 & 10.54 & $-$4.7 & 37.5 & M3 & 8.5 & $<$40 & $-$4.2 & -- \\
50 & 04224274--6137581 & 15.91 & 10.85 & $-$4.6 & 31.7 & M3.5 & 11.1 & $<$40 & $-$5.0 & Broad CCF \\
51 & 03302115--6412207 & 16.59 & 11.29 & 11.9 & 23.5 & M4 & 8.1 & $<$40 & $-$8.5 & Broad CCF \\
52 & 02200131--5909599 & 11.63 & 8.72 & 19.5 & 30.8 & K4 & 1.2 & 10 $\pm$ 10 & 0.3 & -- \\
53 & 05561480--6100328 & 15.69 & 10.81 & $-$12.4 & 23.6 & M3 & 3.6 & $<$40 & $-$5.0 & -- \\
54 & 05584222--5639409 & 15.61 & 11.02 & $-$14.1 & 17.3 & M2.5 & 13.4 & $<$40 & $-$4.0 & -- \\
55 & 05015289--6101308 & 15.37 & 10.65 & $-$5.4 & 25.8 & M3 & 5.4 & $<$40 & $-$5.0 & Broad CCF \\
56 & 05434439--5223259 & 17.39 & 12.30 & $-$4.0 & 21.7 & M4 & 18.9 & $<$40 & $-$4.7 & -- \\
57 & 05425262--4632446 & 14.48 & 10.75 & $-$6.5 & 19.2 & K8 & 17.6 & 235 $\pm$ 10 & $-$1.1 & -- \\
58 & 05184173--4907129 & 13.97 & 10.41 & $-$9.7 & 17.1 & K7 & 12.4 & $<$50 & $-$1.3 & -- \\
59 & 05060048--4545196 & 16.27 & 11.15 & $-$7.0 & 20.5 & M3 & 13.1 & $<$30 & $-$7.5 & -- \\
60 & 03471242--5411160 & 14.41 & 10.10 & 6.6 & 26.2 & M1 & $-$0.3 & $<$40 & $-$1.4 & -- \\
61 & 04081065--4723248 & 12.60 & 9.30 & $-$10.2 & 21.1 & K5 & 17.3 & 30 $\pm$ 20 & $-$0.6 & -- \\
\hline
\end{tabular}
\end{table*}

%% file: table_previousmembers.tex
\begin{table*}
\caption{ Octans members proposed in the literature. $V$ magnitudes and proper motions are from SPM4, except for CD$-$30 3394AB (both  from UCAC4), CD$-$72 248, ($V$ from UCAC4), TYC 7066-1037-1 ($V$ from UCAC4, proper motion from Tycho-2), TYC 9300-0529-1 and TYC 9300-0891-1 (proper motions from PPMXL). Membership and radial velocity references: T03: \protect\cite{Torres03}, T06: \protect\cite{Torres06}, T08: \protect\cite{Torres08}, M11: \protect\cite{Moor11}, M13: \protect\cite{Murphy13}, E14: \protect\cite{Elliott14}, CT: Carlos A. O. Torres, private communication. Li\,\textsc{i} $\lambda$6708 equivalent widths are taken from \protect\citet{Torres06}, except for \rxjstar\ (M13) and BD$-$20 1111 (CT). The T08 members CD$-$72 248 and CD$-$66 395 are fast rotators ($v\sin i=190$~\kms) with assumed large velocity errors.  \label{tab:previousmembers}}
\begin{tabular}{lcccccccccccc}
\hline
Name & Ref. & $V$ & $K_{s}$ & $\mu_{\alpha}\cos\delta$ & $\mu_{\delta}$ & Spectral & RV & Ref.$_{\textrm{RV}}$ & Li EW & Remarks \\
 &  & (mag) & (mag) & (mas yr$^{-1}$) & (mas yr$^{-1}$) & Type & (km s$^{-1}$) &  & (m\AA) &  \\
\hline
CD$-$58 860 & T08 & 9.92 & 8.36 & $-$3.0 & 30.0 & G6 & 9.5 $\pm$ 0.1 & E14 & 225 & -- \\
CD$-$43 1451 & T08 & 10.64 & 8.73 & 2.6 & 19.6 & G9 & 13.2 $\pm$ 0.4 & E14 & 280 & -- \\
CD$-$72 248 & T03 & 10.87 & 8.67 & $-$6.4 & 20.8 & K0 & 4.0 & T06 & 350 & Fast rot. \\
HD 274576 & T08 & 10.53 & 8.81 & $-$11.5 & 19.0 & G6 & 12.1 $\pm$ 0.7 & E14 & 235 & -- \\
BD$-$20 1111 & E14 & 10.41 & 8.70 & $-$8.9 & 10.0 & G7 & 18.7 $\pm$ 1.1 & CT & 200 & -- \\
HD 36968 & M11 & 8.98 & 8.11 & $-$4.3 & 15.7 & F2 & 15.0 $\pm$ 2.0 & M11 & -- & -- \\
CD$-$47 1999 & T08 & 10.05 & 8.64 & $-$8.2 & 15.4 & G0 & 14.4 $\pm$ 1.0 & T06 & 190 & -- \\
TYC 7066-1037-1 & T08 & 11.19 & 9.38 & $-$12.7 & 13.0 & G9 & 14.5 $\pm$ 0.3 & E14 & 250 & -- \\
CD$-$49 2037 & E14 & 11.34 & 9.08 & $-$10.8 & 17.3 & K0 & 12.1 $\pm$ 0.2 & E14 & 290 & -- \\
CD$-$66 395 & T08 & 10.77 & 9.00 & $-$11.1 & 21.5 & K0 & 8.0 & T06 & 250 & Fast rot. \\
CD$-$30 3394A & T08 & 9.89 & 8.59 & $-$16.1 & 10.3 & F6 & 14.5 $\pm$ 0.2 & E14 & 120 & -- \\
CD$-$30 3394B & T08 & 9.97 & 8.70 & $-$13.4 & 9.3 & F9 & 14.9 $\pm$ 1.1 & E14 & 170 & -- \\
RX J1123.2$-$7924 & M13 & 13.71 & 9.67 & $-$30.6 & $-$16.8 & M1.5 & 2.7 $\pm$ 2.9 & M13 & 130 & SB1 \\
HD 155177 & T08 & 8.86 & 7.62 & 11.3 & $-$22.7 & F6 & 0.3 $\pm$ 1.0 & CT & 70 & SB2 \\
TYC 9300-0529-1 & T03 & 11.53 & 9.52 & 14.1 & $-$16.4 & K0 & $-$2.3 $\pm$ 0.5 & E14 & 300 & -- \\
TYC 9300-0891-1 & T03 & 10.81 & 8.73 & 14.8 & $-$16.1 & K0 & $-$3.0 $\pm$ 1.0 & T06 & 300 & 0.9$^{\prime\prime}$ dbl. \\
CP$-$79 1037 & T03 & 11.26 & 9.28 & 16.7 & $-$14.1 & G8 & $-$1.2 $\pm$ 0.4 & T06 & 257 & -- \\
CP$-$82 784 & T03 & 10.90 & 8.63 & 23.3 & $-$13.1 & K1 & $-$2.2 $\pm$ 0.9 & E14 & 275 & -- \\
CD$-$87 121 & T03 & 10.03 & 8.03 & 26.6 & 15.2 & G8 & $-$0.9 $\pm$ 1.4 & E14 & 266 & -- \\
\hline
\end{tabular}
\end{table*}

%% file: table_members.tex
\begin{table*}
\caption{Octans members selected by the kinematic analysis in \S\ref{sec:finaluvw}. Kinematic distances ($d_{\textrm{kin}}$), heliocentric velocities ($U,V,W$) and positions ($X,Y,Z$) were computed from the mean space motion (bottom row) and the proper motions and radial velocities in Tables~\ref{tab:spectroscopy} and \ref{tab:previousmembers}. Also listed are the space motion ($\Delta v$), radial velocity ($\Delta$RV) and total proper motion ($\Delta|\mu|$) residuals. All of the stars selected as kinematic members satisfy $\Delta v<5$~\kms. \label{tab:members}}
\begin{tabular}{lcccccccccccc}
\hline
Name & $d_{\textrm{kin}}$ & $M_{V}$ & $\Delta|\mu|$ & $\Delta$RV & $\Delta v$ & $U$ & $V$ & $W$ & $X$ & $Y$ & $Z$ \\
 & (pc) & (mag) & (mas yr$^{-1}$) & (km s$^{-1}$) & (km s$^{-1}$) & (km s$^{-1}$) & (km s$^{-1}$) & (km s$^{-1}$) & (pc) & (pc) & (pc) \\
\hline
CD$-$58 860 & 99 & 4.94 & 3.8 & $-$1.7 & 2.5 & $-$12.8 & $-$2.5 & $-$10.9 & $-$1.5 & $-$71.7 & $-$68.2 \\
CD$-$43 1451 & 113 & 5.37 & 4.7 & $-$1.3 & 2.8 & $-$13.7 & $-$5.6 & $-$8.2 & $-$31.5 & $-$75.9 & $-$77.6 \\
CD$-$72 248 & 161 & 4.83 & 1.3 & $-$3.2 & 3.4 & $-$13.9 & $-$1.6 & $-$9.9 & 32.1 & $-$130.2 & $-$89.1 \\
HD 274576 & 118 & 5.17 & 1.7 & $-$1.1 & 1.5 & $-$12.9 & $-$3.5 & $-$11.1 & $-$29.5 & $-$94.3 & $-$64.5 \\
CD$-$47 1999 & 156 & 4.08 & 0.9 & 1.7 & 1.8 & $-$14.4 & $-$6.4 & $-$11.2 & $-$36.1 & $-$129.0 & $-$79.9 \\
TYC 7066-1037-1 & 127 & 5.67 & 0.2 & 0.0 & 0.1 & $-$13.7 & $-$4.8 & $-$10.8 & $-$55.6 & $-$100.3 & $-$54.5 \\
CD$-$66 395 & 141 & 5.02 & 3.7 & 0.1 & 2.5 & $-$14.5 & $-$6.0 & $-$9.0 & 14.1 & $-$124.5 & $-$64.7 \\
CD$-$30 3394A & 132 & 4.29 & 1.4 & 0.9 & 1.2 & $-$14.6 & $-$5.5 & $-$10.5 & $-$64.2 & $-$109.7 & $-$35.6 \\
CD$-$30 3394B & 155 & 4.02 & 1.8 & 1.3 & 1.8 & $-$15.1 & $-$5.8 & $-$10.2 & $-$75.4 & $-$128.8 & $-$41.8 \\
HD 155177 & 150 & 2.98 & 2.3 & $-$0.5 & 1.7 & $-$13.3 & $-$3.3 & $-$11.7 & 81.3 & $-$107.7 & $-$65.6 \\
TYC 9300-0529-1 & 175 & 5.31 & 0.8 & 0.2 & 0.7 & $-$13.3 & $-$4.4 & $-$11.4 & 126.1 & $-$95.0 & $-$75.5 \\
TYC 9300-0891-1 & 173 & 4.62 & 1.6 & $-$0.5 & 1.4 & $-$13.7 & $-$3.5 & $-$11.5 & 124.7 & $-$93.9 & $-$74.7 \\
CP$-$79 1037 & 175 & 5.04 & 0.9 & $-$1.0 & 1.3 & $-$14.5 & $-$4.7 & $-$9.9 & 108.4 & $-$107.5 & $-$85.5 \\
CP$-$82 784 & 142 & 5.14 & 3.2 & $-$2.8 & 3.5 & $-$14.5 & $-$1.4 & $-$10.8 & 81.6 & $-$93.9 & $-$68.6 \\
CD$-$87 121 & 123 & 4.58 & 2.6 & $-$3.3 & 3.6 & $-$14.7 & $-$1.4 & $-$10.1 & 59.0 & $-$87.9 & $-$62.6 \\
BD$-$20 1111 & 111 & 5.18 & 3.0 & 2.1 & 2.6 & $-$15.9 & $-$6.1 & $-$10.5 & $-$71.7 & $-$69.0 & $-$49.1 \\
CD$-$49 2037 & 140 & 5.61 & 1.7 & 0.1 & 1.1 & $-$14.2 & $-$5.3 & $-$10.1 & $-$28.6 & $-$120.6 & $-$65.2 \\
RX J1123.2$-$7924 & 107 & 8.56 & 7.1 & 1.7 & 4.0 & $-$10.8 & $-$4.3 & $-$13.6 & 49.2 & $-$89.6 & $-$31.8 \\
HD 36968 & 137 & 3.30 & 5.0 & 0.6 & 3.3 & $-$14.9 & $-$6.7 & $-$8.5 & $-$50.3 & $-$105.7 & $-$71.1 \\
1 & 141 & 8.83 & 2.1 & 0.4 & 1.5 & $-$14.4 & $-$5.7 & $-$10.0 & $-$47.0 & $-$110.2 & $-$74.4 \\
2 & 143 & 10.77 & 0.2 & $-$0.8 & 0.8 & $-$13.5 & $-$4.2 & $-$10.4 & $-$49.6 & $-$116.5 & $-$66.4 \\
4 & 86 & 9.86 & 0.1 & 0.7 & 0.7 & $-$13.9 & $-$5.2 & $-$11.4 & $-$37.3 & $-$57.7 & $-$51.8 \\
8 & 111 & 8.36 & 5.1 & 3.4 & 4.4 & $-$17.0 & $-$7.5 & $-$10.1 & $-$66.6 & $-$74.7 & $-$47.9 \\
9 & 132 & 9.81 & 1.0 & $-$0.2 & 0.7 & $-$14.0 & $-$4.7 & $-$10.3 & $-$41.7 & $-$119.5 & $-$37.6 \\
11 & 85 & 11.65 & 0.1 & 1.8 & 1.8 & $-$14.4 & $-$6.2 & $-$11.7 & $-$32.1 & $-$70.9 & $-$34.1 \\
12 & 116 & 6.97 & 0.2 & 2.9 & 2.9 & $-$14.9 & $-$7.2 & $-$11.9 & $-$47.6 & $-$96.9 & $-$42.4 \\
13 & 110 & 10.74 & 5.1 & 3.0 & 4.0 & $-$16.7 & $-$7.1 & $-$9.7 & $-$63.3 & $-$81.0 & $-$39.2 \\
14 & 134 & 11.45 & 5.2 & 2.1 & 3.9 & $-$12.8 & $-$8.1 & $-$8.9 & 95.6 & $-$85.4 & $-$39.0 \\
15 & 167 & 6.97 & 1.1 & $-$0.2 & 0.9 & $-$14.0 & $-$5.3 & $-$10.3 & 118.7 & $-$101.6 & $-$58.9 \\
23 & 162 & 10.52 & 0.5 & 3.2 & 3.2 & $-$11.6 & $-$6.6 & $-$12.4 & 100.7 & $-$109.7 & $-$63.7 \\
24 & 139 & 8.36 & 2.7 & 3.2 & 3.7 & $-$11.8 & $-$6.8 & $-$13.4 & 42.3 & $-$122.2 & $-$51.1 \\
26 & 130 & 8.80 & 0.2 & $-$1.4 & 1.4 & $-$14.1 & $-$3.7 & $-$10.1 & 40.5 & $-$107.4 & $-$61.1 \\
27 & 126 & 10.40 & 2.3 & $-$2.2 & 2.6 & $-$12.7 & $-$2.4 & $-$10.9 & $-$12.0 & $-$109.4 & $-$61.4 \\
28 & 140 & 9.56 & 0.5 & 1.2 & 1.2 & $-$14.0 & $-$6.0 & $-$11.1 & $-$20.8 & $-$128.1 & $-$52.5 \\
34 & 181 & 8.26 & 2.2 & $-$0.4 & 1.9 & $-$13.7 & $-$3.0 & $-$11.5 & 136.2 & $-$74.1 & $-$93.4 \\
35 & 166 & 8.14 & 2.3 & 0.6 & 1.9 & $-$13.3 & $-$6.5 & $-$10.2 & 124.9 & $-$67.9 & $-$85.7 \\
39 & 157 & 9.76 & 5.0 & 0.3 & 3.7 & $-$12.1 & $-$2.0 & $-$12.9 & 94.6 & $-$93.7 & $-$83.2 \\
40 & 99 & 8.65 & 5.0 & 3.3 & 4.1 & $-$11.2 & $-$4.9 & $-$14.2 & 50.9 & $-$60.1 & $-$60.0 \\
41 & 140 & 8.71 & 0.1 & 1.0 & 1.0 & $-$13.0 & $-$5.3 & $-$11.5 & 102.1 & $-$52.3 & $-$80.3 \\
45 & 129 & 7.69 & 1.7 & $-$0.8 & 1.3 & $-$13.7 & $-$3.6 & $-$11.1 & 57.8 & $-$89.5 & $-$72.7 \\
48 & 89 & 10.45 & 5.4 & 2.5 & 3.4 & $-$12.1 & $-$5.3 & $-$13.9 & 19.1 & $-$67.7 & $-$54.5 \\
49 & 89 & 10.72 & 5.9 & 0.3 & 2.5 & $-$12.4 & $-$3.5 & $-$12.7 & 15.2 & $-$67.5 & $-$55.9 \\
50 & 98 & 10.95 & 3.6 & 0.8 & 1.9 & $-$12.9 & $-$4.4 & $-$12.6 & 3.6 & $-$73.8 & $-$64.4 \\
51 & 121 & 11.18 & 6.2 & $-$1.3 & 3.8 & $-$14.0 & $-$6.4 & $-$7.5 & 14.4 & $-$84.2 & $-$85.7 \\
54 & 138 & 9.91 & 2.9 & 2.8 & 3.4 & $-$12.7 & $-$6.6 & $-$13.7 & $-$10.5 & $-$119.6 & $-$68.0 \\
55 & 120 & 9.97 & 0.8 & $-$4.9 & 4.9 & $-$13.9 & $-$1.1 & $-$7.7 & 0.8 & $-$96.2 & $-$71.7 \\
58 & 136 & 8.30 & 2.6 & $-$0.4 & 1.7 & $-$12.6 & $-$3.8 & $-$11.9 & $-$27.6 & $-$107.9 & $-$78.1 \\
59 & 116 & 10.95 & 0.0 & $-$0.6 & 0.6 & $-$13.5 & $-$4.3 & $-$10.6 & $-$29.6 & $-$87.7 & $-$69.9 \\
\hline \textbf{Mean} ($N=48$) & \textbf{132} & & \textbf{2.5} & \textbf{0.3} & \textbf{2.3} & \textbf{$-$13.7} & \textbf{$-$4.8} & \textbf{$-$10.9}\\ \hline
\end{tabular}
\end{table*}

%% file: table_possible.tex
\begin{table*}
\caption{Possible Octans members not included in the final membership solution. The absolute magnitude given for the components of \#32AB assumes an equal-mass system with combined brightness $M_{V}=5.77$~mag. Notes for last column: ($\Delta v$):  Rejected as kinematic member with $\Delta v>5$~\kms, (H$\alpha$): H$\alpha$ in absorption but otherwise acceptable kinematic member, (Li): Lithium-rich star with EW$_{\textrm{Li I}}>100$~m\AA.\label{tab:possible}}
\begin{tabular}{lcccccccccccc}
\hline
ID & $d_{\textrm{kin}}$ & $M_{V}$ & $\Delta|\mu|$ & $\Delta$RV & $\Delta v$ & $U$ & $V$ & $W$ & $X$ & $Y$ & $Z$ & Remarks \\
 & (pc) & (mag) & (mas yr$^{-1}$) & (km s$^{-1}$) & (km s$^{-1}$) & (km s$^{-1}$) & (km s$^{-1}$) & (km s$^{-1}$) & (pc) & (pc) & (pc) &  \\
\hline
16 & 110 & 7.08 & 1.5 & $-$3.8 & 3.8 & $-$16.2 & $-$2.3 & $-$9.5 & 65.1 & $-$84.9 & $-$25.7 & H$\alpha$ \\
32A & 175 & 5.02 & 1.1 & 6.9 & 6.9 & $-$8.3 & $-$7.2 & $-$14.6 & 131.9 & $-$80.9 & $-$81.8 & Li, $\Delta v$ \\
32B & 175 & 5.02 & 1.1 & 2.7 & 2.8 & $-$11.5 & $-$5.3 & $-$12.7 & 131.9 & $-$80.9 & $-$81.8 & H$\alpha$ \\
36 & 163 & 8.02 & 4.9 & $-$5.7 & 6.8 & $-$16.6 & 1.3 & $-$9.7 & 116.1 & $-$82.4 & $-$79.3 & Li, $\Delta v$ \\
57 & 129 & 8.93 & 4.6 & 4.6 & 5.4 & $-$15.7 & $-$9.8 & $-$10.9 & $-$32.3 & $-$106.0 & $-$66.0 & Li, $\Delta v$ \\
\hline
\end{tabular}
\end{table*}

%% file: table_rosat.tex
\begin{table*}
\caption{ Candidates with \emph{ROSAT} counterparts lying on or above the RASS limit in Fig.~\ref{fig:rosat}. Stars already identified with \emph{ROSAT} sources in SACY \citep{Torres06,Torres08,Elliott14} or earlier works are not included. Stars \#25/26 and \#34/35 are matched to the same sources. The 0.1--2.4 keV X-ray luminosity ($L_{X}$) was derived from the kinematic distance, count rate and hardness ratio using the conversion factor given in \protect\cite{Fleming95b}, while fractional X-ray luminosities ($\log L_{X}/L_{\textrm{bol}}$) were calculated from the $V-K_{s}$ colour, $V$ magnitude and the bolometric corrections tabled by \citet{Pecaut13}. \label{tab:rosat}}
\begin{tabular}{lccccccccc}
\hline
ID & Survey & 1RXS & Offset & Error & Count Rate & HR1 & $\log L_{X}$ & $\log L_{X}/L_{\textrm{bol}}$ & Member? \\
 &  &  & (arcsec) & (arcsec) & (cnt s$^{-1}$) &  & (erg s$^{-1}$) &  &  \\
\hline
12 & BSC & 1RXS J062045.9$-$362003 & 22 & 12 & 0.081 $\pm$ 0.013 & $-$0.4 $\pm$ 0.13 & 29.9 & $-$3.11 & Y \\
61 & FSC & 1RXS J040810.0$-$472357 & 34 & 29 & 0.068 $\pm$ 0.023 & 0.05 $\pm$ 0.31 & 29.8 & $-$2.97 &  \\
35 & FSC & 1RXS J195446.1$-$645101 & 70 & 24 & 0.049 $\pm$ 0.021 & $-$0.16 $\pm$ 0.44 & 30.1 & $-$2.85 & Y \\
34 & FSC & 1RXS J195446.1$-$645101 & 57 & 24 & 0.049 $\pm$ 0.021 & $-$0.16 $\pm$ 0.44 & 30.2 & $-$2.67 & Y \\
48 & FSC & 1RXS J040852.1$-$720357 & 9 & 14 & 0.043 $\pm$ 0.011 & $-$0.42 $\pm$ 0.22 & 29.4 & $-$2.90 & Y \\
44 & FSC & 1RXS J045750.1$-$842950 & 22 & 13 & 0.035 $\pm$ 0.008 & 0.56 $\pm$ 0.23 & 29.9 & $-$3.18 &  \\
8 & FSC & 1RXS J054107.5$-$240907 & 11 & 12 & 0.030 $\pm$ 0.010 & $-$0.19 $\pm$ 0.3 & 29.5 & $-$3.03 & Y \\
1 & FSC & 1RXS J053219.0$-$413233 & 32 & 14 & 0.029 $\pm$ 0.009 & $-$0.25 $\pm$ 0.26 & 29.7 & $-$2.86 & Y \\
47 & FSC & 1RXS J044732.0$-$741202 & 30 & 15 & 0.028 $\pm$ 0.010 & 0.81 $\pm$ 0.35 & 29.9 & $-$2.86 &  \\
2 & FSC & 1RXS J055412.9$-$404353 & 13 & 15 & 0.026 $\pm$ 0.007 & 0.47 $\pm$ 0.27 & 29.8 & $-$2.21 & Y \\
26 & FSC & 1RXS J062109.5$-$790446 & 64 & 42 & 0.018 $\pm$ 0.007 & 0.47 $\pm$ 0.42 & 29.6 & $-$3.12 & Y \\
25 & FSC & 1RXS J062109.5$-$790446 & 64 & 42 & 0.018 $\pm$ 0.007 & 0.47 $\pm$ 0.42 & 29.5 & $-$5.42 &  \\
\hline
\end{tabular}
\end{table*}